\documentclass[aps,prd,twocolumn,superscriptaddress,showpacs,floatfix]{revtex4}
\usepackage{dcolumn}
\usepackage{natbib}
\usepackage{amsmath}
\usepackage{amssymb}
\usepackage{subfigure}
\usepackage{xspace}
\usepackage{color}
\usepackage{footnote}
\usepackage{longtable}

\usepackage{graphicx}
\DeclareGraphicsExtensions{.eps,.ps,.eps.gz,.ps.gz}
\newcommand{\href}[2]{#2}                   
\usepackage{lscape}

\graphicspath{{figs/}}

\def\goes{\rightarrow}

\newcommand{\tzc}{$t\rightarrow Zc$}
\newcommand{\zee}{$Z\rightarrow e^+e^-$}
\newcommand{\mmet}{{\,\rm\not E}_{\rm T}}

\newcommand{\Et}{E_{\rm T}}

\newcommand{\Pt}{{\rm p}_{\rm T}}

\newcommand{\wenu}{$W\rightarrow e\nu$}
\newcommand{\wmunu}{$W\rightarrow\mu\nu$}
\newcommand{\zmumu}{$Z\rightarrow\mu^+\mu^-$}

\newcommand{\ppbar}{$p\overline{p}$}
\newcommand{\ttbar}{$t\overline{t}$}

\newcommand{\lum}{1.52 $\mathrm{fb^{-1}}$}
\newcommand{\invpb}{$\mathrm{fb^{-1}}$}
\newcommand{\degs}{\mbox{$^{\circ}$}}

\newcommand{\met}{\mbox{${\rm \not\! E}_{\rm T}$}}

\usepackage{feynmp}
\newcommand{\plus}{\kern -0.18em +\kern -0.18em}

\newcommand{\MeV}{\ensuremath{\mathrm{\ Me\kern -0.1em V}}\xspace}
\newcommand{\MeVc}{\ensuremath{\mathrm{\ Me\kern -0.1em V\kern -0.1em  \mathit{c}}}\xspace}
\newcommand{\MeVcsq}{\ensuremath{\mathrm{\ Me\kern -0.1em V\kern -0.1em \mathit{c}^2}}\xspace}
\newcommand{\GeV}{\ensuremath{\mathrm{Ge\kern -0.1em V}}\xspace}
\newcommand{\GeVc}{\ensuremath{\mathrm{\ Ge\kern -0.1em V\kern -0.1em \mathit{c}}}\xspace}
\newcommand{\GeVcsq}{\ensuremath{\mathrm{\ Ge\kern -0.1em V\kern -0.1em \mathit{c}^2}}\xspace}
\newcommand{\TeV}{\ensuremath{\mathrm{Te\kern -0.1em V}}\xspace}
\newcommand{\bfTeV}{\ensuremath{\bf{Te\kern -0.1em V}}\xspace}
\newcommand{\nsGeV}{\ensuremath{\mathrm{Ge\kern -0.1em V}}\xspace}
\newcommand{\nsGeVc}{\ensuremath{\mathrm{Ge\kern -0.1em V\kern -0.1em \mathit{c}}}\xspace}

\begin{document}
%
\title{Search for the Neutral Current Top Quark Decay $t\goes Zc$ 
Using the Ratio of $Z$-Boson + 4 Jets to $W$-Boson + 4 Jets Production}


\affiliation{Institute of Physics, Academia Sinica, Taipei, Taiwan 11529, Republic of China} 
\affiliation{Argonne National Laboratory, Argonne, Illinois 60439} 
\affiliation{University of Athens, 157 71 Athens, Greece} 
\affiliation{Institut de Fisica d'Altes Energies, Universitat Autonoma de Barcelona, E-08193, Bellaterra (Barcelona), Spain} 
\affiliation{Baylor University, Waco, Texas  76798} 
\affiliation{Istituto Nazionale di Fisica Nucleare Bologna, $^x$University of Bologna, I-40127 Bologna, Italy} 
\affiliation{Brandeis University, Waltham, Massachusetts 02254} 
\affiliation{University of California, Davis, Davis, California  95616} 
\affiliation{University of California, Los Angeles, Los Angeles, California  90024} 
\affiliation{University of California, San Diego, La Jolla, California  92093} 
\affiliation{University of California, Santa Barbara, Santa Barbara, California 93106} 
\affiliation{Instituto de Fisica de Cantabria, CSIC-University of Cantabria, 39005 Santander, Spain} 
\affiliation{Carnegie Mellon University, Pittsburgh, PA  15213} 
\affiliation{Enrico Fermi Institute, University of Chicago, Chicago, Illinois 60637}
\affiliation{Comenius University, 842 48 Bratislava, Slovakia; Institute of Experimental Physics, 040 01 Kosice, Slovakia} 
\affiliation{Joint Institute for Nuclear Research, RU-141980 Dubna, Russia} 
\affiliation{Duke University, Durham, North Carolina  27708} 
\affiliation{Fermi National Accelerator Laboratory, Batavia, Illinois 60510} 
\affiliation{University of Florida, Gainesville, Florida  32611} 
\affiliation{Laboratori Nazionali di Frascati, Istituto Nazionale di Fisica Nucleare, I-00044 Frascati, Italy} 
\affiliation{University of Geneva, CH-1211 Geneva 4, Switzerland} 
\affiliation{Glasgow University, Glasgow G12 8QQ, United Kingdom} 
\affiliation{Harvard University, Cambridge, Massachusetts 02138} 
\affiliation{Division of High Energy Physics, Department of Physics, University of Helsinki and Helsinki Institute of Physics, FIN-00014, Helsinki, Finland} 
\affiliation{University of Illinois, Urbana, Illinois 61801} 
\affiliation{The Johns Hopkins University, Baltimore, Maryland 21218} 
\affiliation{Institut f\"{u}r Experimentelle Kernphysik, Universit\"{a}t Karlsruhe, 76128 Karlsruhe, Germany} 
\affiliation{Center for High Energy Physics: Kyungpook National University, Daegu 702-701, Korea; Seoul National University, Seoul 151-742, Korea; Sungkyunkwan University, Suwon 440-746, Korea; Korea Institute of Science and Technology Information, Daejeon, 305-806, Korea; Chonnam National University, Gwangju, 500-757, Korea} 
\affiliation{Ernest Orlando Lawrence Berkeley National Laboratory, Berkeley, California 94720} 
\affiliation{University of Liverpool, Liverpool L69 7ZE, United Kingdom} 
\affiliation{University College London, London WC1E 6BT, United Kingdom} 
\affiliation{Centro de Investigaciones Energeticas Medioambientales y Tecnologicas, E-28040 Madrid, Spain} 
\affiliation{Massachusetts Institute of Technology, Cambridge, Massachusetts  02139} 
\affiliation{Institute of Particle Physics: McGill University, Montr\'{e}al, Qu\'{e}bec, Canada H3A~2T8; Simon Fraser University, Burnaby, British Columbia, Canada V5A~1S6; University of Toronto, Toronto, Ontario, Canada M5S~1A7; and TRIUMF, Vancouver, British Columbia, Canada V6T~2A3} 
\affiliation{University of Michigan, Ann Arbor, Michigan 48109} 
\affiliation{Michigan State University, East Lansing, Michigan  48824}
\affiliation{Institution for Theoretical and Experimental Physics, ITEP, Moscow 117259, Russia} 
\affiliation{University of New Mexico, Albuquerque, New Mexico 87131} 
\affiliation{Northwestern University, Evanston, Illinois  60208} 
\affiliation{The Ohio State University, Columbus, Ohio  43210} 
\affiliation{Okayama University, Okayama 700-8530, Japan} 
\affiliation{Osaka City University, Osaka 588, Japan} 
\affiliation{University of Oxford, Oxford OX1 3RH, United Kingdom} 
\affiliation{Istituto Nazionale di Fisica Nucleare, Sezione di Padova-Trento, $^y$University of Padova, I-35131 Padova, Italy} 
\affiliation{LPNHE, Universite Pierre et Marie Curie/IN2P3-CNRS, UMR7585, Paris, F-75252 France} 
\affiliation{University of Pennsylvania, Philadelphia, Pennsylvania 19104}
\affiliation{Istituto Nazionale di Fisica Nucleare Pisa, $^z$University of Pisa, $^{aa}$University of Siena and $^{bb}$Scuola Normale Superiore, I-56127 Pisa, Italy} 
\affiliation{University of Pittsburgh, Pittsburgh, Pennsylvania 15260} 
\affiliation{Purdue University, West Lafayette, Indiana 47907} 
\affiliation{University of Rochester, Rochester, New York 14627} 
\affiliation{The Rockefeller University, New York, New York 10021} 
\affiliation{Istituto Nazionale di Fisica Nucleare, Sezione di Roma 1, $^{cc}$Sapienza Universit\`{a} di Roma, I-00185 Roma, Italy} 

\affiliation{Rutgers University, Piscataway, New Jersey 08855} 
\affiliation{Texas A\&M University, College Station, Texas 77843} 
\affiliation{Istituto Nazionale di Fisica Nucleare Trieste/Udine, I-34100 Trieste, $^{dd}$University of Trieste/Udine, I-33100 Udine, Italy} 
\affiliation{University of Tsukuba, Tsukuba, Ibaraki 305, Japan} 
\affiliation{Tufts University, Medford, Massachusetts 02155} 
\affiliation{Waseda University, Tokyo 169, Japan} 
\affiliation{Wayne State University, Detroit, Michigan  48201} 
\affiliation{University of Wisconsin, Madison, Wisconsin 53706} 
\affiliation{Yale University, New Haven, Connecticut 06520} 
\author{T.~Aaltonen}
\affiliation{Division of High Energy Physics, Department of Physics, University of Helsinki and Helsinki Institute of Physics, FIN-00014, Helsinki, Finland}
\author{J.~Adelman}
\affiliation{Enrico Fermi Institute, University of Chicago, Chicago, Illinois 60637}
\author{T.~Akimoto}
\affiliation{University of Tsukuba, Tsukuba, Ibaraki 305, Japan}
\author{B.~\'{A}lvarez~Gonz\'{a}lez$^s$}
\affiliation{Instituto de Fisica de Cantabria, CSIC-University of Cantabria, 39005 Santander, Spain}
\author{S.~Amerio$^y$}
\affiliation{Istituto Nazionale di Fisica Nucleare, Sezione di Padova-Trento, $^y$University of Padova, I-35131 Padova, Italy} 

\author{D.~Amidei}
\affiliation{University of Michigan, Ann Arbor, Michigan 48109}
\author{A.~Anastassov}
\affiliation{Northwestern University, Evanston, Illinois  60208}
\author{A.~Annovi}
\affiliation{Laboratori Nazionali di Frascati, Istituto Nazionale di Fisica Nucleare, I-00044 Frascati, Italy}
\author{J.~Antos}
\affiliation{Comenius University, 842 48 Bratislava, Slovakia; Institute of Experimental Physics, 040 01 Kosice, Slovakia}
\author{G.~Apollinari}
\affiliation{Fermi National Accelerator Laboratory, Batavia, Illinois 60510}
% ...

\collaboration{CDF Collaboration\footnote{With visitors from $^a$University of Massachusetts Amherst, Amherst, Massachusetts 01003,
$^b$Universiteit Antwerpen, B-2610 Antwerp, Belgium, 
$^c$University of Bristol, Bristol BS8 1TL, United Kingdom,
$^d$Chinese Academy of Sciences, Beijing 100864, China, 
$^e$Istituto Nazionale di Fisica Nucleare, Sezione di Cagliari, 09042 Monserrato (Cagliari), Italy,
$^f$University of California Irvine, Irvine, CA  92697, 
$^g$University of California Santa Cruz, Santa Cruz, CA  95064, 
$^h$Cornell University, Ithaca, NY  14853, 
$^i$University of Cyprus, Nicosia CY-1678, Cyprus, 
$^j$University College Dublin, Dublin 4, Ireland,
$^k$University of Edinburgh, Edinburgh EH9 3JZ, United Kingdom, 
$^l$University of Fukui, Fukui City, Fukui Prefecture, Japan 910-0017
$^m$Kinki University, Higashi-Osaka City, Japan 577-8502
$^n$Universidad Iberoamericana, Mexico D.F., Mexico,
$^o$Queen Mary, University of London, London, E1 4NS, England,
$^p$University of Manchester, Manchester M13 9PL, England, 
$^q$Nagasaki Institute of Applied Science, Nagasaki, Japan, 
$^r$University of Notre Dame, Notre Dame, IN 46556,
$^s$University de Oviedo, E-33007 Oviedo, Spain, 
$^t$Texas Tech University, Lubbock, TX  79609, 
$^u$IFIC(CSIC-Universitat de Valencia), 46071 Valencia, Spain,
$^v$University of Virginia, Charlottesville, VA  22904,
$^w$Bergische Universit\"at Wuppertal, 42097 Wuppertal, Germany,
$^{ee}$On leave from J.~Stefan Institute, Ljubljana, Slovenia, 
}}
\noaffiliation

%
\date{\today}
\begin{abstract}

We have used the Collider Detector at Fermilab (CDF II) to search for
the flavor-changing neutral-current (FCNC) top quark decay ${t\goes
Zc}$ using a technique employing ratios of $W$ and $Z$ production,
measured in \ppbar~data corresponding to an integrated luminosity of
\lum. The analysis uses a comparison of two decay chains,
${p\bar{p}\goes t\bar{t}\goes WbWb\goes\ell\nu bjjb}$ and
${p\bar{p}\goes t\bar{t}\goes ZcWb\goes\ell\ell cjjb}$, to cancel
systematic uncertainties in acceptance, efficiency, and luminosity. We
validate the modeling of acceptance and efficiency for lepton
identification over the multi-year dataset using another ratio of $W$
and $Z$ production, in this case the observed ratio of inclusive
production of $W$ to $Z$ bosons. To improve the discrimination against
standard model backgrounds to top quark decays, we calculate the top
quark mass for each event with two leptons and four jets assuming it
is a $t\bar{t}$ event with one of the top quarks decaying to $Zc$. For
additional background discrimination we require at least one jet to be
identified as originating from a $b$-quark. No significant signal is
found and we set an upper limit on the FCNC branching ratio
$Br({t\goes Zc})$ using a likelihood constructed from the ${\ell\ell
cjjb}$ top quark mass distribution and the number of ${\ell\nu bjjb}$
events. Limits are set as a function of the helicity of the $Z$ boson
produced in the FCNC decay. For 100\% longitudinally polarized
$Z$ bosons we find limits of 8.3\% and 9.3\% (95\% C.L.) depending on
the assumptions regarding the theoretical top quark pair production
cross-section.

\end{abstract}


\maketitle

\section{Introduction}

\par The standard model (SM) Lagrangian does not contain any
flavor-changing neutral current (FCNC) terms such as ${d\goes s}$, a
consequence of its SU(2) structure \citep{t_May_FCNC}. In the SM the
top quark is expected to decay via the charged weak current into a $W$
boson and a bottom quark, ${t\goes W^+b}$, with close to 100\%
branching ratio~\citep{PDG}. We test this prediction by searching for
FCNC interactions in top quark decays, in which the top quark decays
to a $Z$ boson and a charm quark, ${t\goes Zc}$. In the standard model
the FCNC decay ${t\goes Zc}$ is highly suppressed, proceeding only
through radiative corrections, with a predicted branching ratio
$Br({t\goes Zc})$ of about $10^{-14}$ \citep{FCNC_Expect}. However,
some extensions of the SM (e.g. two-Higgs doublet models, models with
extra quark singlets, technicolor models with a dynamical breakdown of
the electroweak symmetry, etc.) predict measurable
rates~\citep{t_May_FCNC,Ewk_FCNC_Constrain,FCNC_top}.

\par The production of top quark pairs, \ttbar, is the preferred
channel to observe the FCNC transition ${t\goes c}$ at the Tevatron,
as single top quark production has a smaller cross-section and much
larger QCD backgrounds in the $Zc$ final state. We have used data from
an integrated luminosity of \lum~collected with the CDF II
detector~\citep{CDFdetector} at the Fermilab Tevatron to search for
events in which one of the top quarks decays to $Zc$ and the other one
decays to $Wb$.  In order to get a sample of high purity, we select
the leptonic decays of the $Z$ boson, \zee~and \zmumu. In this
scenario, the FCNC signature is most likely a pair of
oppositely-charged leptons forming a $Z$ boson, and four jets (the $b$
and $c$ jets from the $t$ and ${\bar{t}}$, and two jets from ${W\goes
q\bar{q}'}$), with the event being kinematically consistent with the
FCNC \ttbar~decay hypothesis. We require at least one jet with a
displaced secondary vertex as a sign of a heavy-flavor quark ($b$ or
$c$-quark) to further suppress hadronic backgrounds.

\par
\begin{figure}[h]
\centering
\includegraphics[angle=0,width=0.40\textwidth]{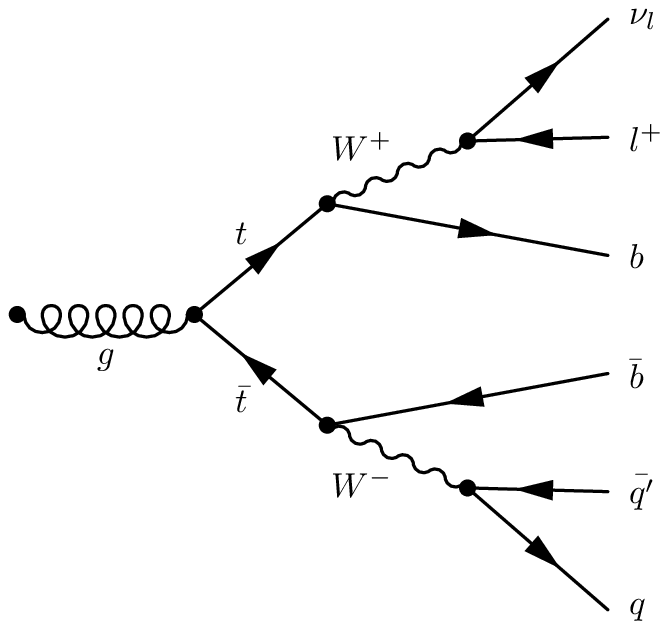}
\caption{A Feynman diagram for one of the processes contributing to 
${p\bar{p}\goes t\bar{t}\goes WbWb\goes\ell\nu bjjb}$ decay chain.}
\label{fig:fdiag_WbWb}
\end{figure}

\begin{figure}[h]
\centering
\includegraphics[angle=0,width=0.40\textwidth]{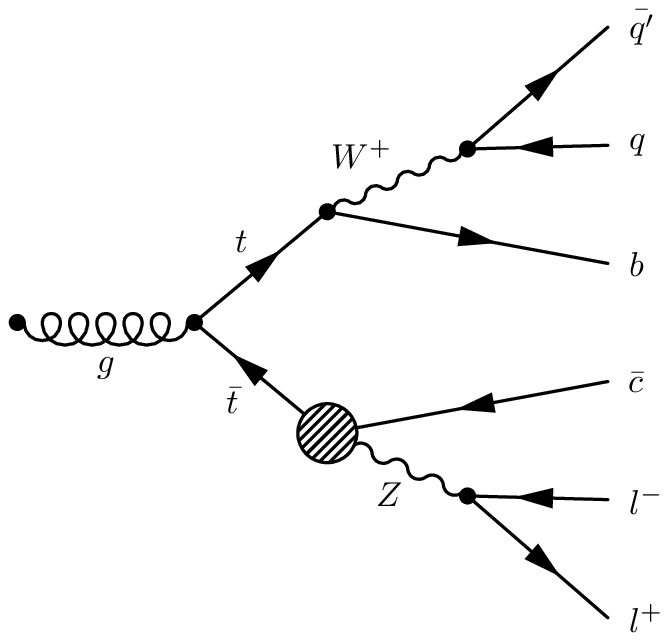}
\caption{A Feynman diagram for one of the processes contributing to 
${p\bar{p}\goes t\bar{t}\goes ZcWb\goes\ell\ell cjjb}$ decay chain.
The blob represents a non-SM FCNC vertex.  }
\label{fig:fdiag_WbZc}
\end{figure}

 To minimize the systematic uncertainties on the particle
identification and trigger efficiencies, geometric acceptances, and
luminosity, we rely on a technique based on the simultaneous
comparison of two decay chains:
\begin{enumerate}
\setlength{\itemsep}{-0.03in}
\item ${p\bar{p}\goes t\bar{t}\goes WbWb\goes \ell\nu bjjb}$ (see 
Fig.~\ref{fig:fdiag_WbWb}),
\item ${p\bar{p}\goes t\bar{t}\goes ZcWb\goes\ell\ell cjjb}$ (see 
Fig.~\ref{fig:fdiag_WbZc}).
\end{enumerate}
Many of the systematic uncertainties contributing to both decay chains
are correlated and tend to cancel, improving the precision and
robustness of the result. The other decay modes of top quark pairs
(e.g. ${t\bar{t}\goes WbWb\goes\ell_1\nu_1 b\ell_2\nu_2b}$,
${t\bar{t}\goes WbWb\goes jjbjjb}$, or ${t\bar{t}\goes
ZcWb\goes\ell_1^+\ell_1^-c\ell_2\nu_2b}$) have low acceptances or high
levels of background and are not used.
\par The final states $\ell\nu bjjb$ and $\ell\ell cjjb$ used
in this analysis contain products of the leptonic decays of
$W\goes\ell\nu$ and $Z\goes\ell\ell$, for which there exist precise
next-to-next-to-leading order (NNLO) predictions of inclusive
cross-sections multiplied by branching fractions~\citep{R_ratio}. We
use a comparison of the measured ratio of inclusive $W\goes\ell\nu$ to
$Z\goes\ell\ell$ production to validate the lepton identification and
trigger efficiencies in the Monte Carlo simulation predictions of
signal and SM background to about 2\%. This technique, which parallels
that used in the search, will be used for precision comparisons with
the standard model at the LHC~\citep{erin_prd}.

\par We present a technique for measuring the background in the
inclusive $W$ boson sample coming from QCD multi-jet events using a
data-derived model~\citep{Cooper,Messina}. The number of
mis-identified $W$ bosons is estimated separately for electrons and
muons by fitting the observed distributions in \met~\citep{EtPt} with
templates from real $W$ decays and modeled non-$W$ events.

\par We also present a simple technique to estimate the number of
muons from cosmic rays in the $Z$ boson sample, using the distribution
in the magnitude of the vector momentum sum, $|\vec{P}(\mu^+\mu^-)|$,
of the muon pair. This is an economical way of combining the usual
``back-to-back'' and momentum balance criteria for the two muons into
a single distribution, as ${\mu^+\mu^-}$ pairs from cosmic rays have a
very narrow peak at $|\vec{P}(\mu^+\mu^-)|=0$ GeV, while real
\zmumu~decays occupy a much larger volume in the 3-dimensional momentum
space.

\par A previous limit on the branching ratio for \tzc~\citep{PDG} is 
from CDF using data from Run I of the Tevatron; the limit is 33\% at
95\% confidence level (C.L.) \citep{FCNC_Limit}. The limit from
precision measurements at LEP is lower, 13.7\% at 95\%
C.L.~\citep{LEP_indirect}.

\par There is a recent CDF limit from a parallel independent analysis,
using a different technique and a total luminosity of 1.9 \invpb, of
3.7\% at 95\% C.L., more restrictive than the result presented
here~\citep{CDF_direct}.  The technique presented here is specifically
designed to reduce systematic errors by using ratios of $W$ and $Z$
boson events, important for much larger integrated
luminosities~\cite{erin_prd}.
The acceptance for \ttbar$\goes ZcWb\goes\ell\ell cjjb$ events in this
analysis is 0.303\% versus the 0.43\% quoted in
Ref.~\citep{CDF_direct}.

\par We derive the first upper limits on $Br(t\goes Zc)$ as a function
of the polarization of the $Z$ boson produced in a FCNC decay of a top
quark. These limits cover all possible FCNC top-quark couplings.


\par The outline of the paper is as follows. Section~\ref{detector}
briefly describes the CDF II detector. The analysis strategy, which
uses the SM decay modes of the top quark as well as the signal FCNC
decay mode to allow cancellation of major systematic uncertainties of
acceptance, efficiency, and luminosity, is described in
Section~\ref{analysis}. Section~\ref{data} describes the event
selection, which starts with the dataset of events selected on
central~\citep{CDF_coo} high transverse momentum~\citep{EtPt}
electrons and muons. The identification of jets containing heavy
flavor is given in Section~\ref{btag}.  Sections~\ref{zdef}
and~\ref{wdef} describe the selection of $Z$ and $W$ bosons,
respectively. The modeling and validation of standard model vector
boson production and the estimation of backgrounds are presented in
Section~\ref{datasets}.  Section~\ref{R_ratios} describes the
technique of using the measurement of $R$, the ratio of inclusive $W$
boson to inclusive $Z$ boson production, as a check of the complex
Monte Carlo samples generated using the detector and accelerator
conditions accumulated over the long period of data taking.
Section~\ref{FCNC_Modeling} describes the modeling of the FCNC signal
from top quark decay.  The measurement of the expected SM
contributions to the signature $W$ boson + 4-jets with one jet
identified as heavy-flavor, dominated by top quark pair production, is
described in Section~\ref{ttbar}. The contributions to the reference
channel, $W$ boson + 4-jets, and the signal channel, $Z$ boson +
4-jets, each with one or more heavy-flavor jets, are presented in
Sections~\ref{FCNC_Wjets} and~\ref{FCNC_Zjets}, respectively.

\par The estimation of systematic uncertainties on the acceptances and
backgrounds is described in Section~\ref{systematics}, and the limit
calculations for a full range of possible longitudinal FCNC couplings
are presented in Section~\ref{final_limit}. Section~\ref{conclusions}
is a summary of the conclusions.

\section{The CDF II Detector}
\label{detector}

\par The CDF II detector is a cylindrically symmetric spectrometer
designed to study \ppbar~collisions at the Fermilab Tevatron. The
detector has been extensively described in the
literature~\citep{CDFdetector}. Here we briefly describe the detector
subsystems relevant for the analysis.

\par Tracking systems are used to measure the momenta of charged
particles, and to trigger on and identify leptons with large
transverse momentum, $\Pt$~\citep{EtPt}. A multi-layer system of
silicon strip detectors~\citep{SVX}, which identifies tracks in both
the $r-\phi$ and $r-z$ views~\citep{CDF_coo}, and the central outer
tracker (COT)~\citep{COT} are contained in a superconducting solenoid
that generates a magnetic field of 1.4 T. The COT is a 3.1 m long
open-cell drift chamber that makes up to 96 measurements along the
track of each charged particle in the region $|\eta|<1$. Sense wires
are arranged in 8 alternating axial and stereo ($\pm 2\degs$)
super-layers with 12 wires each. For high momentum tracks, the COT
$\Pt$ resolution is $\sigma_{\Pt}/\Pt^2 \simeq
0.0017~\GeV^{-1}$~\citep{CDF_Momentum}.

\par Segmented calorimeters with towers arranged in a projective
geometry, each tower consisting of an electromagnetic and a hadronic
compartment~\citep{cem_resolution, cal_upgrade}, cover the central
region, $|\eta|<1$ (CEM/CHA), and the `end plug' region,
$1<|\eta|<3.6$ (PEM/PHA).  In both the central and end plug regions,
systems with finer spatial resolution are used to make profile
measurements of electromagnetic showers at shower
maximum~\citep{CDFII} for electron identification (the CES and PES
systems, respectively).  Electrons are reconstructed in the CEM with
an $\Et$~\citep{EtPt} resolution of $\sigma(\Et)/\Et \simeq
13.5\%/\sqrt{\Et/\GeV}\oplus 2\%$~\citep{cem_resolution} and in the
PEM with an $\Et$ resolution of $\sigma(\Et)/\Et \simeq
16.0\%/\sqrt{\Et/\GeV}\oplus 1\%$~\citep{pem_resolution}.  Jets are
identified using a cone clustering algorithm in $\eta-\phi$ space of
radius 0.4 as a group of electromagnetic and hadronic calorimeter
towers; the jet energy resolution is approximately $\sigma \simeq
0.1\cdot\Et ({\GeV}) +1.0~\GeV$~\citep{jet_resolution}.

\par Muons are identified using the central CMU, CMP, and
CMX~\citep{cmu_ref, cmp_ref} muon systems, which cover the kinematic
region $|\eta|<1$. The CMU system uses four layers of planar drift
chambers to detect muons with $\Pt>1.4~\GeV$ in the central region of
$|\eta|<0.6$.  The CMP system consists of an additional four layers of
planar drift chambers located behind 0.6 m of steel outside the
magnetic return yoke, and detects muons with $\Pt>2.0~\GeV$. The CMX
detects muons in the region $0.6<|\eta|<1.0$ with four to eight layers
of drift chambers, depending on the polar angle.

\par The beam luminosity is measured using two sets of gas Cherenkov
counters, located in the region $3.7<|\eta|<4.7$. The total
uncertainty on the luminosity is estimated to be 5.9\%, where 4.4\%
comes from the acceptance and operation of the luminosity monitor and
4.0\% from the calculation of the inelastic
\ppbar~cross-section~\citep{luminosity}.

\par A 3-level trigger system~\citep{CDFdetector} selects events for
further analysis offline.  The first two levels of triggers consist of
dedicated fast digital electronics analyzing a subset of the full
detector data.  The third level, applied to the full data from the
detector for those events passing the first two levels, consists of a
farm of computers that reconstruct the data and apply selection
criteria for (typically) several hundred distinct triggers.

\section{Introduction to the Analysis Strategy}
\label{analysis}
\par The measurement of the branching ratio of the $t\goes Zc$ decay mode
is designed to be similar to the measurement of the $R$-ratio
between inclusive cross-section of $W$'s to $Z$'s. The ratio $R$ is
defined as:
\begin{equation}
R=\frac{\sigma(W)\cdot Br(W\goes\ell\nu)}{\sigma(Z)\cdot
  Br(Z\goes\ell\ell)},
\label{R_equation}
\end{equation}
where $\sigma(W)$ and $\sigma(Z)$ are cross-sections of inclusively
produced $W$ and $Z$ bosons. A measurement of the $R$-ratio is itself
a precise test of lepton identification efficiencies, triggering, and
Monte Carlo simulations. A measured $R$-ratio has smaller
uncertainties than $\sigma(W)$ and $\sigma(Z)$ since some of the
uncertainties (e.g. for integrated luminosity) completely cancel
out. This makes $R$ a valuable tool for precise comparisons between
experimental and theoretical predictions for channels involving both
$W$ and $Z$ bosons.

\par We estimate $R$ for electrons and muons separately (see
Section~\ref{R_ratios}) since these particles are identified with
different detector subsystems. The observed numbers are consistent
with the theoretical predictions~\citep{Stirling_R,R_Renton} and the
previous CDF measurement~\citep{R_ratio_exp} (see
Section~\ref{R_ratios}). This cross-check was performed before
measuring the branching ratio $Br(t\goes Zc)$.

\par The measurement of $Br(t\goes Zc)$ is designed to be a
measurement of the ratio between events in exclusive final states with
a $Z$ boson and four jets and a $W$ boson and four jets. In the case
of the FCNC scenario, a larger FCNC branching fraction leads to fewer
top quark events decaying to $W$ + 4 jets, and consequently an
increase in the rate of $Z$ + 4 jets events.
We subtract SM non-\ttbar~events from events with a $W$ or a $Z$ boson
and four jets so that the ratio is more sensitive to the FCNC signal.
The ratio of $Z$ + 4 jets to $W$ + 4 jets increases in presence of
FCNC events.

\section{Event Selection}
\label{data}
\par The analysis uses events selected by the trigger system that
contain either a central electron with $\Et>18$ GeV or a muon with
$\Pt>18$ GeV~\citep{EtPt}. The electron dataset contains 75.5M events;
the muon dataset contains about 21.2M events. The integrated
luminosity of each dataset is \lum.

\par Both the observed and the simulated events (see
Section~\ref{montecarlo}) are processed through the same selection
criteria to identify electrons and muons, jets, $W$ and $Z$ bosons,
missing transverse energy, and jets containing heavy flavor. Details
of the selection criteria are provided below.

\subsection{Lepton Identification}
\label{lepton_id}

\par We use standard CDF definitions for identification (ID) of
electrons and muons, as described below~\citep{R_ratio_exp}. The same
lepton ID requirements are applied to events from data and Monte Carlo
simulations.

\par The identification and triggering efficiencies for leptons are
different for events in data and Monte Carlo, although they
demonstrate a very similar energy dependence. To eliminate this
inconsistency we follow the standard CDF practice of using correction
factors (``scale factors'') to re-weight the MC events (see
Section~\ref{MC_lept_corrections}).

\par In order to maintain a high efficiency for $Z$ bosons, for which
we require two identified leptons, we define ``tight'' and ``loose''
selection criteria for both electrons and muons, as described below.

\par To reduce backgrounds from the decays of hadrons produced in 
jets, leptons are required to be ``isolated''. The $\Et$ deposited in
the calorimeter towers in a cone in $\eta-\varphi$
space~\citep{CDF_coo} of radius $R=0.4$ around the lepton position is
summed, and the $\Et$ due to the lepton is subtracted. The remaining
$\Et$ is required to be less than 10\% of the lepton $\Et$ for
electrons or $\Pt$ for muons.

\subsubsection{Electron Selection}
\label{electron}
\par An electron candidate passing the tight selection must be central
with $\Et>20$ \GeV, and have: a) a high quality
track~\citep{electron_track_quality} with $\Pt>0.5\cdot\Et$ or $\Pt >
50$ $\GeV$; b) a good transverse shower profile at shower maximum that
matches the extrapolated track position; c) a lateral sharing of
energy in the two calorimeter towers containing the electron shower
consistent with that expected; and d) minimal leakage into the hadron
calorimeter~\citep{hadoem}.
\par Additional central electrons, classified as ``loose'' electrons,
are required to have $\Et>12~\GeV$ and to satisfy the tight central
electron criteria but with a track requirement of $\Pt>10$ $\GeV$
(rather than $0.5\cdot\Et$), and no requirement on a shower maximum
measurement or lateral energy sharing between calorimeter towers.
Electrons in the end-plug calorimeters ($1.2 < |\eta| < 2.5$), also
classified as ``loose'' electrons, are required to have $\Et>
12~\GeV$, minimal leakage into the hadron calorimeter, a track
containing at least 3 hits in the silicon tracking system, and a
shower transverse shape consistent with that expected, with a centroid
close to the extrapolated position of the
track~\citep{wenu_asymmetry_paper}.

\subsubsection{Muon Selection}
\label{muon}

\par A muon candidate passing the tight cuts must have: a) a well
measured track in the COT~\citep{muon_track_quality} with
$\Pt>20~\GeV$; b) energy deposited in the calorimeter consistent with
expectations~\citep{muon_cal_cuts}; c) a muon ``stub''~\citep{stub} in
both the CMU and CMP, or in the CMX, consistent with the extrapolated
COT track~\citep{muon_stub_matching}; and d) a COT track fit
consistent with an outgoing particle from a \ppbar~collision and not
from an incoming cosmic ray~\citep{CRAY}.
\par Additional muons, classified as ``loose'', are required to have
$\Pt>12~\GeV$ and to satisfy the same criteria as for tight muons but
with relaxed COT track quality requirements. Alternatively, for muons
outside the muon system fiducial volume, a loose muon must satisfy the
tight muon criteria and an additional more stringent requirement on
track quality, but the requirement that there be a matching ``stub''
in the muon systems is dropped.

\subsubsection{Corrections due to Modeling of Electrons and Muons in 
the MC Events}
\label{MC_lept_corrections}
\par Following the standard treatment of lepton efficiencies in CDF,
we re-weight Monte Carlo events to take into account the difference
between the identification efficiencies measured in leptonic $Z$
decays and those used in simulation~\citep{lepton_scale_factors}. We
then make additional corrections for the difference in trigger
efficiencies in simulated events and measured in data. Corrections to
trigger efficiencies are typically 4\% for trigger electrons, 8\% for
trigger muons that traverse both the CMU and CMP systems, and 5\% for
muons in the CMX system. The average weight for \zee~events is 0.939;
for \zmumu~events it is 0.891.

\subsection{Jet Identification}
\label{jet_id}
\par Jets are reconstructed using the standard CDF cone-based
clustering algorithm with a cone radius of $R = 0.4$ within
$|\eta|<2.4$~\citep{jet_clu}. The jet energies are corrected for the
$\eta$-dependent response of the calorimeters and for the
luminosity-dependent effect of multiple-\ppbar~interactions. The
simulated calorimeter response for individual hadrons is tuned to
match that in data~\citep{jet_corr}. The raw energy of the jets must
be greater than 8 GeV and the corrected energy is required to be
greater than 15 GeV. Jets that coincide with an identified electron or
photon are removed; i.e. each calorimeter cluster can be associated
with either a jet, an electron, or a photon, which have mutually
exclusive definitions to avoid any ambiguities.
\par There is one case for which the jet energies are corrected to the
parton level rather than to the hadron level. This is done to
calculate the top quark mass in events with a Z boson and four jets
(see Section~\ref{fitting_top}).
%
%
\par High-$\Pt$ photons are not rare in hard-scattering events.
Identifying photons as jets and then correcting them as jets can lead
to mis-reconstructed missing transverse energy and other kinematic
variables, and can be important in an analysis leading to small signal
samples, as in this analysis. Photon candidates are required to have
no matching track with $\Pt>1$ $\GeV$, and at most one track with
$\Pt<1$ $\GeV$, pointing at the calorimeter cluster; good profiles in
both transverse dimensions at shower maximum; and minimal leakage into
the hadron calorimeter~\citep{hadoem}. We require photons to be
``isolated'' in a slightly more restrictive fashion than that for the
leptons: the sum of the $\Pt$ of all tracks in the cone must be less
than $2.0~\GeV+0.005\times\Et$.

\subsection{Reconstruction of Missing Transverse Energy}

\par Missing transverse energy (\met) is the negative two-dimensional
vector sum of $\vec{E}_T$ of all identified objects in the event:
electrons, muons, photons, jets, and unclustered energy. The
unclustered energy is calculated as a two-dimensional vector of raw
calorimeter energy corrected for the energy deposited by identified
jets, electrons, muons, and photons. 


\subsection{Tagging of Heavy Flavor Jets}
\label{btag}

\par We identify decays of bottom and charm quarks (heavy flavor, HF)
with an algorithm that identifies displaced secondary vertexes within
a jet. The primary vertex is identified by fitting all prompt tracks
in the event to a vertex constrained to lie on the beam-line.  Jets
with $E_T$ $>$ 15 GeV are checked for good quality tracks with hits in
the COT and the silicon detector. At least two good tracks consistent
with a common vertex are required to form a secondary vertex
candidate. The distance is calculated between the primary vertex and
the secondary vertex candidate and projected on the jet direction.
The jet is considered to contain a HF quark (``$b$-tagged'') if the
significance of this distance is greater than 7.5$\sigma$. The
algorithm has an efficiency of approximately 50\% to tag a $b$-jet,
depending on the $E_T$ of the jet, in a $t\bar{t}$ event. More details
of the algorithm are available in~\citep{SECVTX}.
\par To model the multiple SM sources of tagged events, we use control
samples selected from the data to estimate mis-tag rates (i.e. the
fraction of tags coming from non-HF-jets), and Monte Carlo simulated
samples to get the contribution from SM physics processes with true
heavy flavor jets.
\par The contribution from real HF jets is estimated by applying the
tagging algorithm to $Z$+HF and $W$+HF MC samples. Events with at
least one $b$-tag are selected. Each selected event is re-weighted by
$(1-(1-\epsilon_{tag})^{N_{\mathrm{tags}}})$ using per-tagged-jet
scale factor
$\epsilon_{tag}$~=~0.95$\pm$0.05~\citep{mistag_rate,SECVTX}, where
$N_{\mathrm{tags}}$ is the number of $b$-tagged jets in the event, to
take into account the difference in the tagging efficiencies between
data and simulation.
\par The mis-tag rate is estimated by applying the mis-tag
parametrization~\citep{mistag_rate} to each event in a data sample
that has all the desired characteristics except a $b$-tag (called the
``pre-tag'' sample). The parametrization gives each jet a probability
to be falsely tagged based on the jet $E_T$, $\eta$, and number of
tracks of good quality in the jet.
\par The calculation of the mis-tag rate is performed in three
steps. First we select all jets with $E_T>10$ GeV and $|\eta|<2.4$ in
the event. We then apply the mis-tag parameterization to the selected
jets. Finally we loop through jets satisfying the event selection
requirements (see Section~\ref{jet_id}) to calculate the probability
for each jet to be falsely tagged as originating from a decay of a
bottom or charm quark.  The per-jet mis-tag probability is roughly 1\%.

\section{Production of $Z$ Bosons with Jets}
\label{zdef}

\par
\begin{figure}[h]
\centering
\includegraphics[angle=0,width=0.45\textwidth]{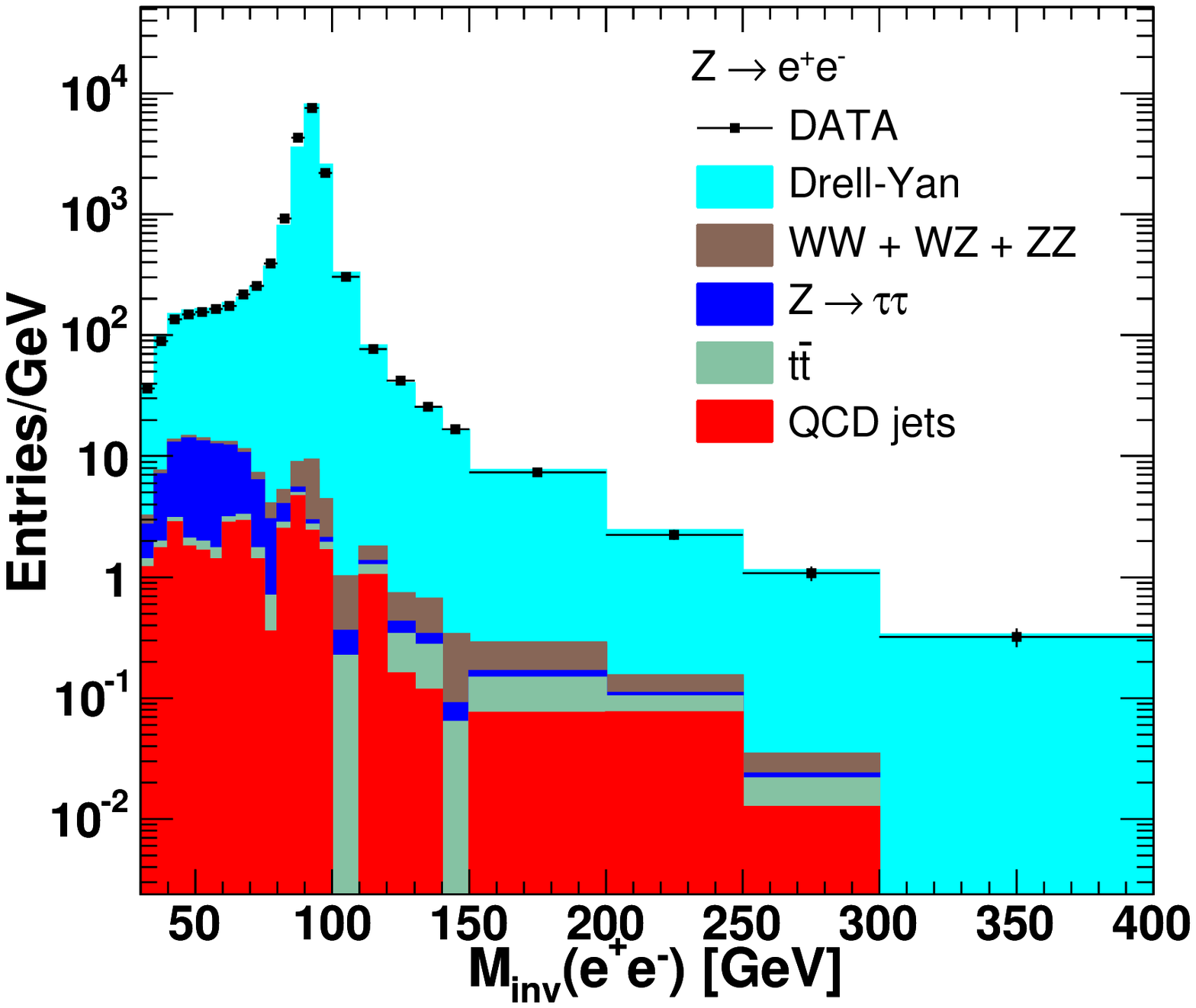}
\includegraphics[angle=0,width=0.45\textwidth]{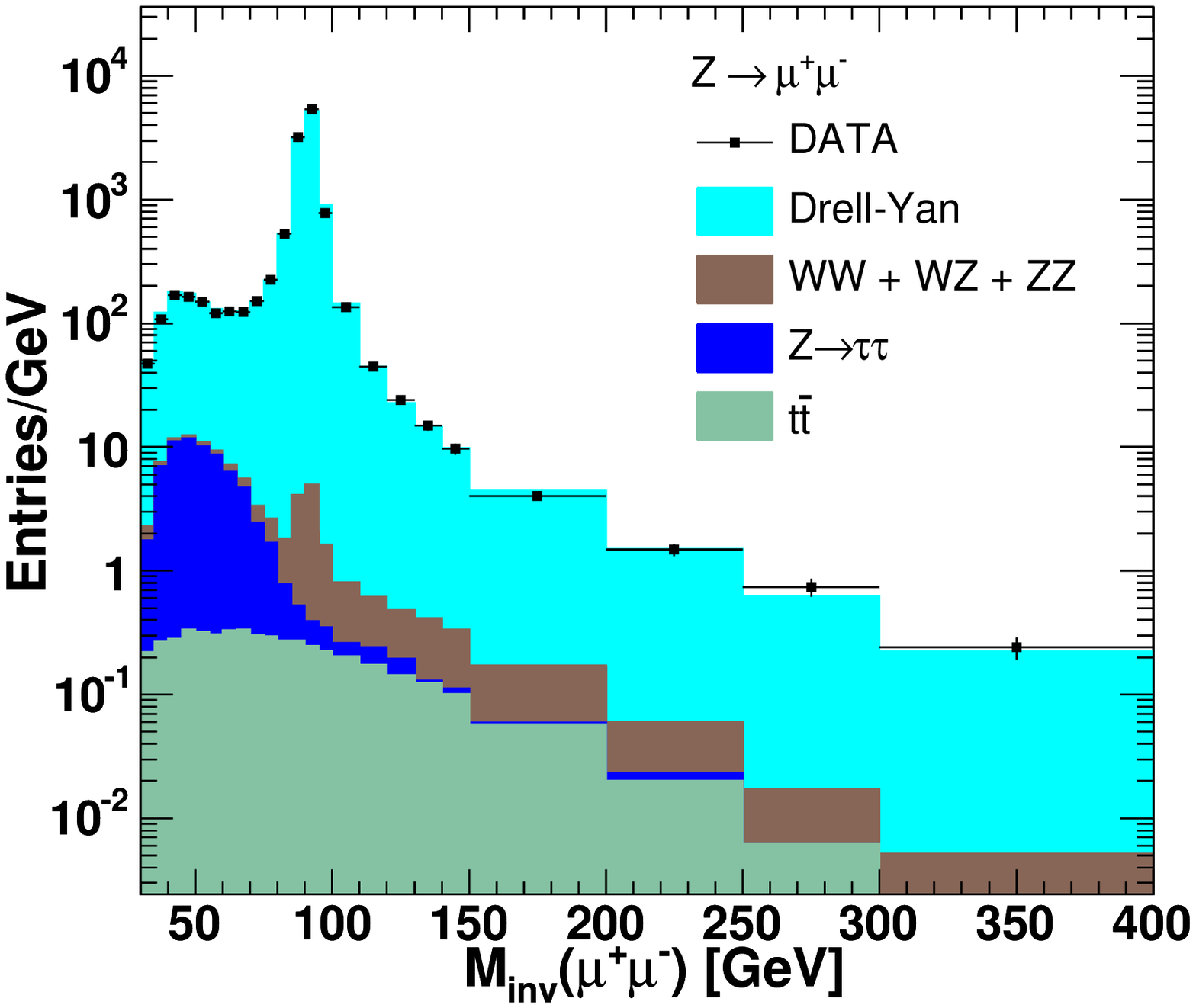}
\caption{The observed (points) and expected (histogram) distributions
in the invariant mass of $e^+e^-$ (upper figure) and $\mu^+\mu^-$
(lower figure) lepton pairs. The order of stacking in the histograms
is the same as in their legends.}
\label{fig:z_m}
\end{figure}

 To be identified as a $Z$ boson a pair of opposite-sign electrons
or muons must have a reconstructed invariant mass in the mass window
from 66 GeV to 116 GeV.  The selection of $Z\goes\ell\ell$ events
requires two tight leptons or a tight and a loose lepton. The two
leptons are required to be assigned the same primary vertex. Figure
\ref{fig:z_m} shows the distributions in invariant mass for electron
and muon pairs.

\par The SM expectation for events with a $Z$ boson and jets is
constructed using Monte Carlo simulations of SM electroweak processes
such as production of $WW$, $WZ$, $ZZ$, and $Z\goes \tau\tau$ (see
Section~\ref{datasets}).

\par The detection of $Z$ bosons is less sensitive to the lepton 
trigger efficiencies than the detection of $W$ bosons, since there are
two leptons in each $Z$ event.


\section{Production of $W$ Bosons with Jets}
\label{wdef}
\par
\begin{figure}[h]
\centering
\includegraphics[angle=0,width=0.45\textwidth]{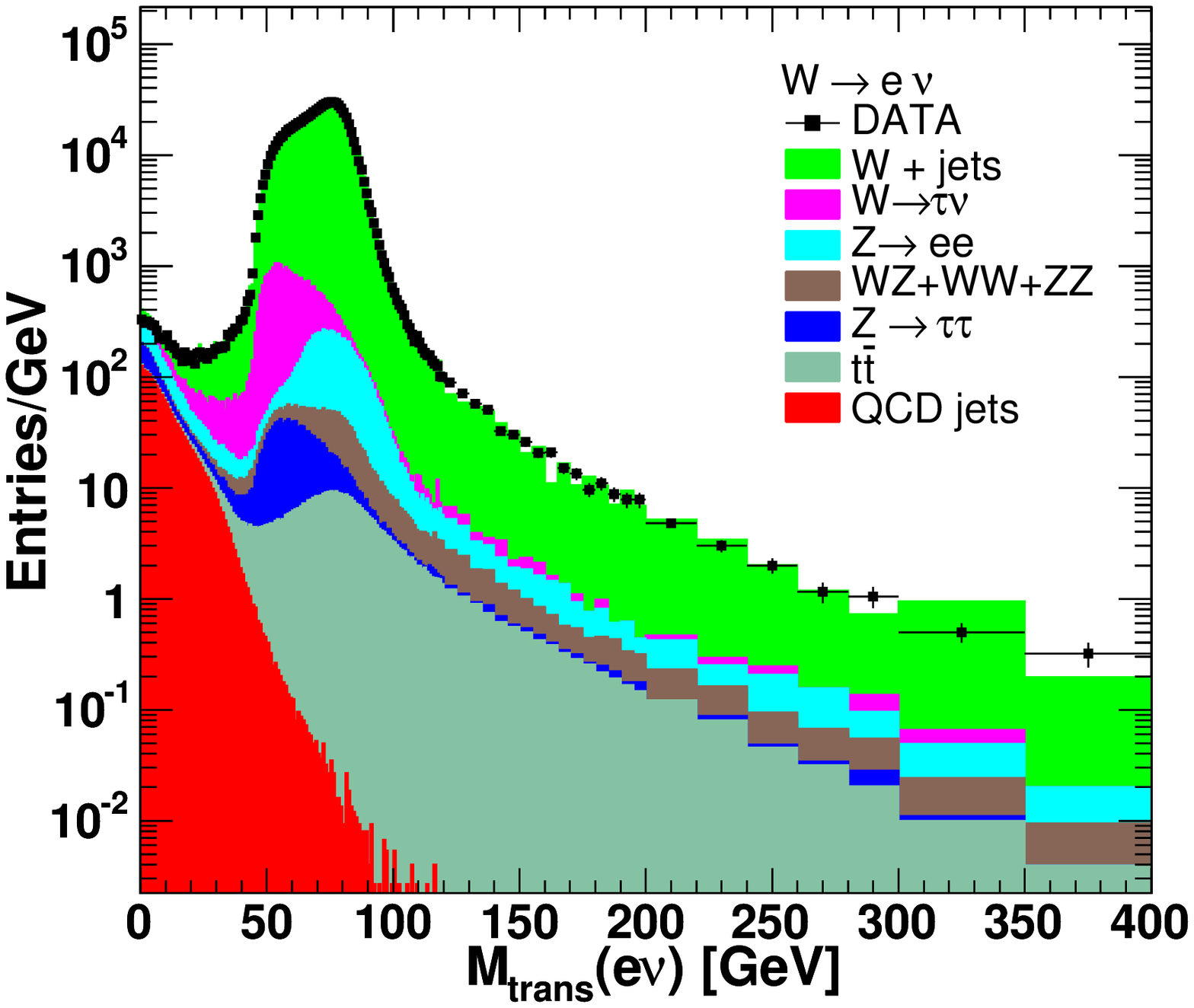}
\includegraphics[angle=0,width=0.45\textwidth]{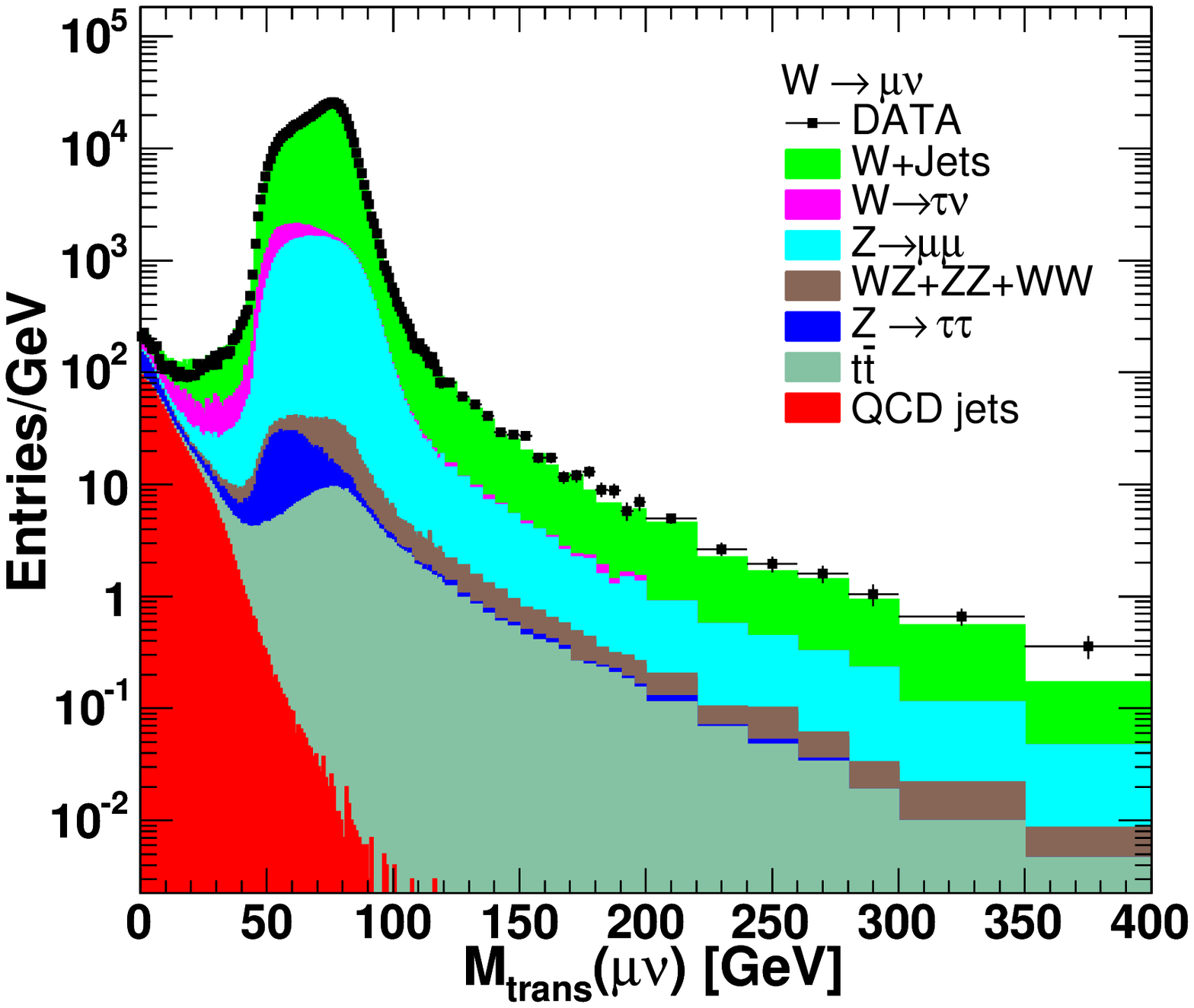}
\caption{The observed (points) and expected (histogram) distributions
in transverse mass of $e+\met$ (upper figure) and $\mu+\met$ (lower
figure). The contribution from $t\bar{t}$ production is calculated for
the case when the top quarks are decaying in the standard way. The
order of stacking in the histograms is the same as in their legends.}
\label{fig:w_mt}
\end{figure}

 The selection of $W\goes\ell\nu$ events requires a tight
central electron or a tight muon and \met~greater than 25 GeV.  We
require that each $W$ event has only one tight lepton, and no loose
leptons. The transverse mass~\citep{transverse_mass},
$M_{\mathrm{trans}}(\ell\nu)$, reconstructed from the lepton and the
missing transverse energy is required to be greater than 20
GeV. Figure~\ref{fig:w_mt} shows the measured and expected
distributions in transverse mass for the
\wenu~and \wmunu~events.

\par The SM backgrounds to events with $W$ + jets (where
$W\goes\ell\nu$) are estimated using the data and from MC
simulations. The MC simulations are used to predict well-understood SM
electroweak processes, such as $Z\goes\ell\ell$, $WW$, $WZ$, and
$ZZ$. Backgrounds that are largely instrumental, such as the
misidentification of a QCD jet as a lepton from $W$ decay, are
predicted from the data. More details are provided in
Section~\ref{datasets}.

\section{Standard Model Contributions to Events with a $W$ or a $Z$ 
Boson and Jets}
\label{datasets}

\subsection{Monte Carlo Simulations of the Standard Model Processes}
\label{montecarlo}
\par The standard model expectations for the production of $W$ and $Z$
bosons are calculated from Monte Carlo simulations. We use {\sc
pythia} to generate $W$ + light jets and $Z$ + light jets processes
and {\sc alpgen} for generation of the heavy flavor processes $W$ + HF
jets and $Z$ + HF jets.
\par The datasets for the $W$ and $Z$ + light jets signatures are
produced using a customized version of {\sc pythia} in which the $\Pt$
spectrum of the $Z$ bosons, $\Pt^Z$, has been tuned to CDF Run I data
for $0 < \Pt^Z < 20$ GeV, and which incorporates a tuned
underlying-event~\citep{Rick_Field} and a requirement that
$M_{\mathrm{inv}}(\ell\ell)$ $>$ 30 GeV. The $W$ and $Z$ + heavy
flavor jets samples are produced with a version of {\sc alpgen} that
has built-in matching of the number of jets from showering and
matrix-element production~\citep{alpgen_matching}.  Showering and
hadronization of jets is done with {\sc
pythia}~\citep{monte_carlo_details}. Events from the MC generators,
{\sc alpgen} and {\sc pythia}, are processed through the full detector
simulation to be reconstructed and analyzed like data.
\par We use the CDF version of {\sc pythia} to describe the inclusive
(i.e. before $b$-tagging) production of $W$ and $Z$ bosons, used for
background calculations. We use primarily {\sc alpgen} samples to
analyze $b$-tagged events, as {\sc pythia} does not handle
heavy-flavor production correctly. {\sc alpgen} handles radiation of
additional jets better than {\sc pythia}. The inclusive production of
$W$ and $Z$ bosons has only a second-order dependence on the
difference in the jet radiation of {\sc alpgen} and {\sc pythia}. It
has a stronger dependence on the momentum distribution of the bosons
which was tuned in the CDF version of {\sc pythia} as described above.
%
%
\par The MC contributions from the SM leading order processes are
combined into inclusive samples using weights proportional to the
cross-sections of each contribution. These summed MC-samples are then
compared to the observed events in the electron and muon decay modes
of $W$ and $Z$ bosons separately. We use NNLO cross-sections of 2.687 
nb and 251.3 pb for the production of $W$ and $Z$ bosons, 
respectively~\citep{R_ratio_exp}.
%

\subsection{Electroweak Backgrounds}
\par Several SM processes other than Drell-Yan production of $W$'s and
$Z$'s contribute to the $W$ and $Z$ leptonic signatures we use in the
analysis, in particular $Z\goes\tau^+\tau^-$, $WW$, $WZ$, $ZZ$,
$W\goes\tau\nu$, and \ttbar $\goes WbWb$. These processes are
estimated from corresponding MC samples, generated using {\sc pythia}.
We weight $WW$, $WZ$, and $ZZ$ datasets using NLO cross-sections (13.0
pb, 3.96 pb, and 1.56 pb, respectively~\citep{Diboson_NLO}).

\subsection{Fake $Z$ Background from Jets Misidentified as Leptons}
\label{z_qcd}

\par This background consists of events in which one or more leptons
are ``fake'', i.e. jets misidentified as leptons. We assume that in
the samples with a vector boson and two-or-more jets, the true lepton
and the fake lepton making up the $Z$ in the background events have no
charge correlation. As the number of fake $Z$ bosons is small (see
below), we use the number of same sign lepton pairs to estimate the
QCD jet background in the $\gamma^*/Z\goes\ell\ell$ sample.
\par The \zmumu~sample, which requires 66 GeV $<$
$M_{\mathrm{inv}}(\ell\ell)$ $<$ 116 GeV, contains only 8 events with
muons of the same sign out of 53,358 total events in the sample. The
fake muon background is consequently negligibly small.

\par Same-sign electron pairs have a significant source from $e^+e^-$
pair-production by photon conversions. The observed number of
same-sign electron pairs in the \zee~sample is corrected for the
predicted number of $e^+e^-$ pairs mis-reconstructed as $e^+e^+$ or
$e^-e^-$ using MC predictions for \zee~production.

\par We observe 398 same-sign electron pairs and 82,901 $e^+e^-$
pairs. We remove the contribution of real $\gamma^*/Z\goes e^+e^-$
events from the number of observed events by subtracting the number of
observed $e^+e^-$ events scaled by the fraction of same-sign to
opposite-sign events in the Monte-Carlo samples for \zee. The
remaining 78 same-sign electron pairs are used to estimate the QCD jet
background in the \zee~sample (see Fig.~\ref{fig:z_m}).

\subsection{Non-$W$ Backgrounds from Jets}
\label{w_qcd}

\par Jet production, which has a much higher cross-section than $W$ or
$Z$ boson production, produces events which mimic the leptonic decay
of a $W$ boson by a mis-measured jet ``faking'' a tight isolated
lepton and large missing energy (\met).
\par
\begin{figure*}[p]
\centering
\includegraphics[angle=0,width=0.45\textwidth]{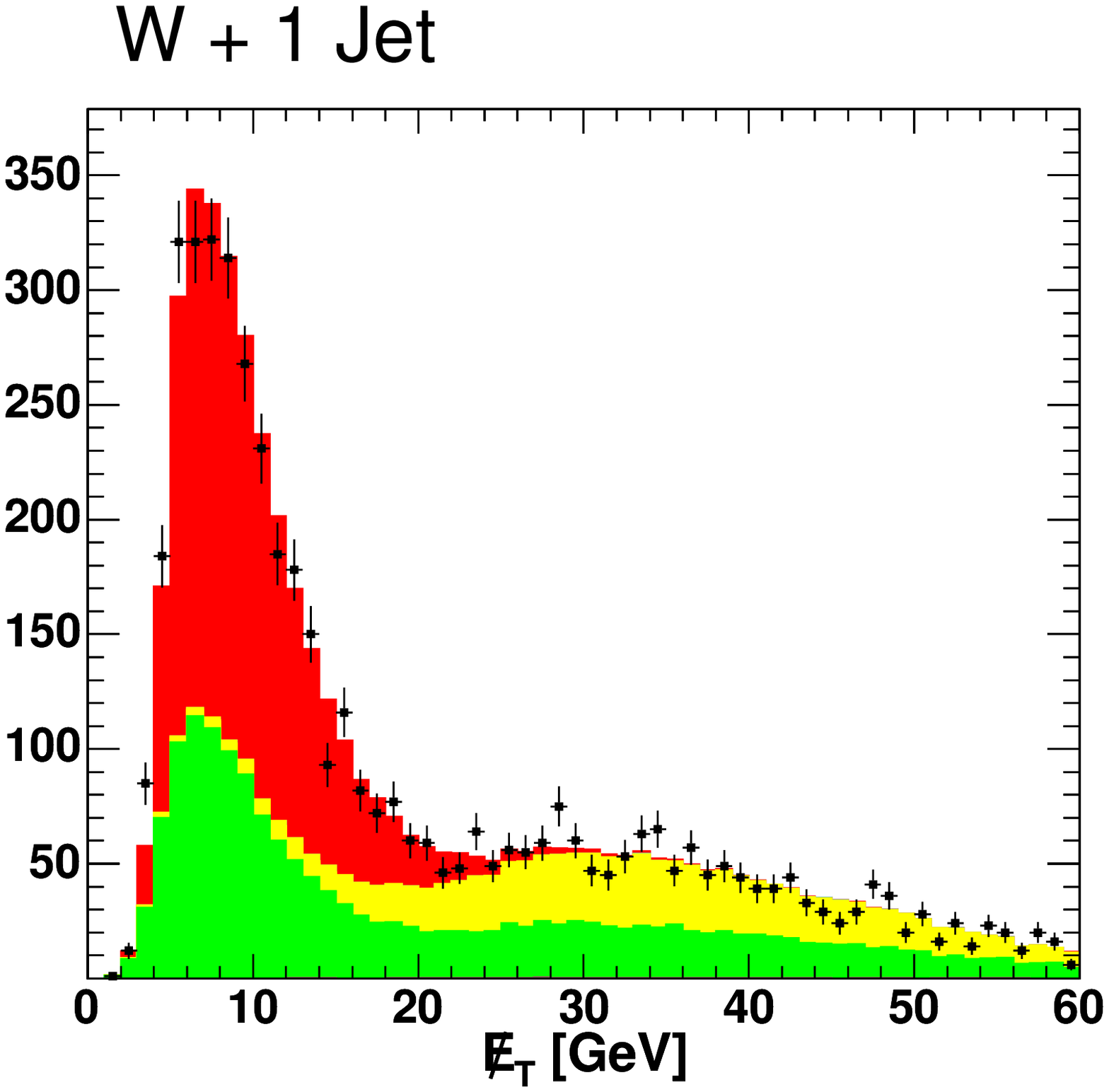}
\includegraphics[angle=0,width=0.45\textwidth]{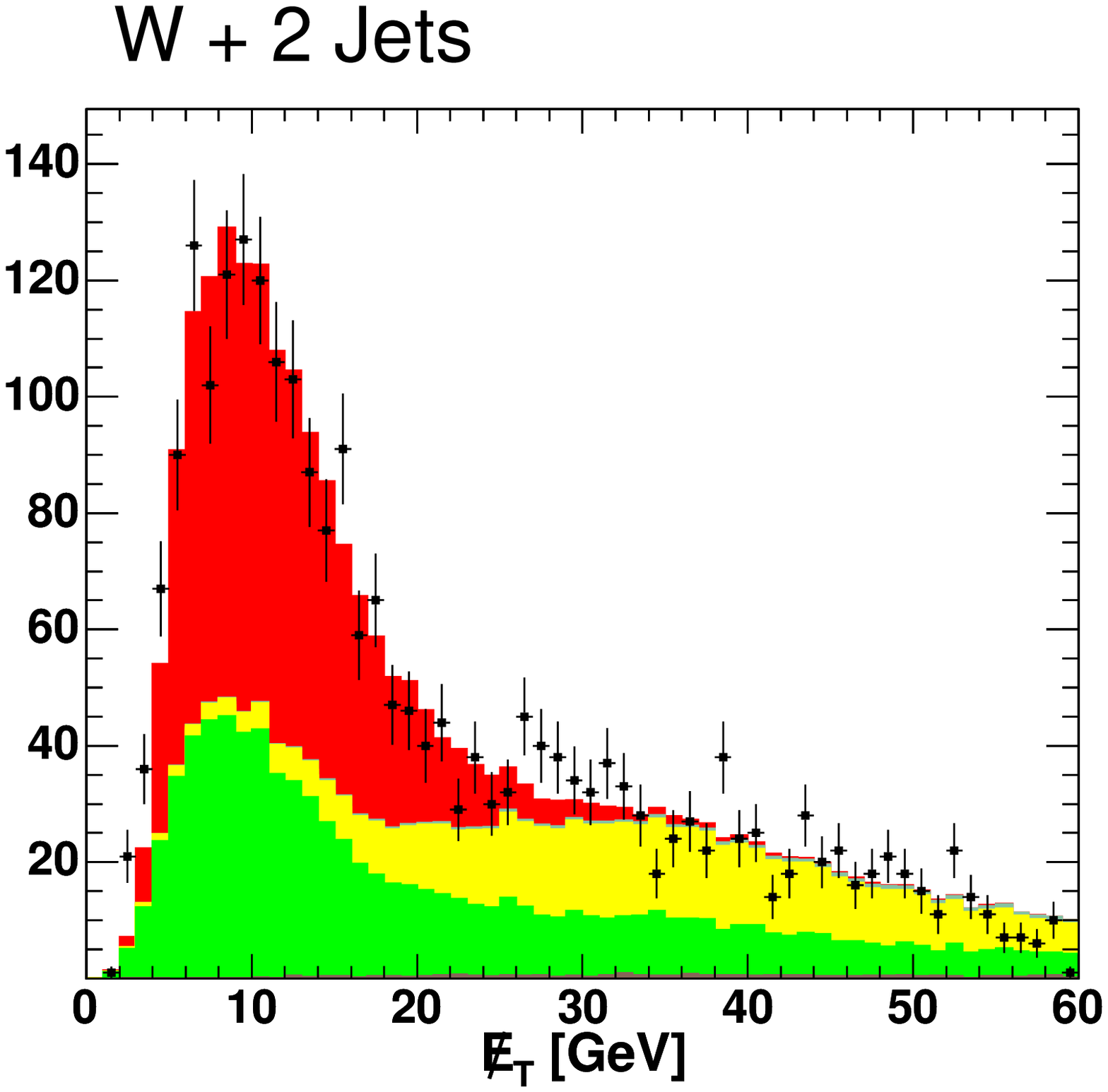}
\includegraphics[angle=0,width=0.45\textwidth]{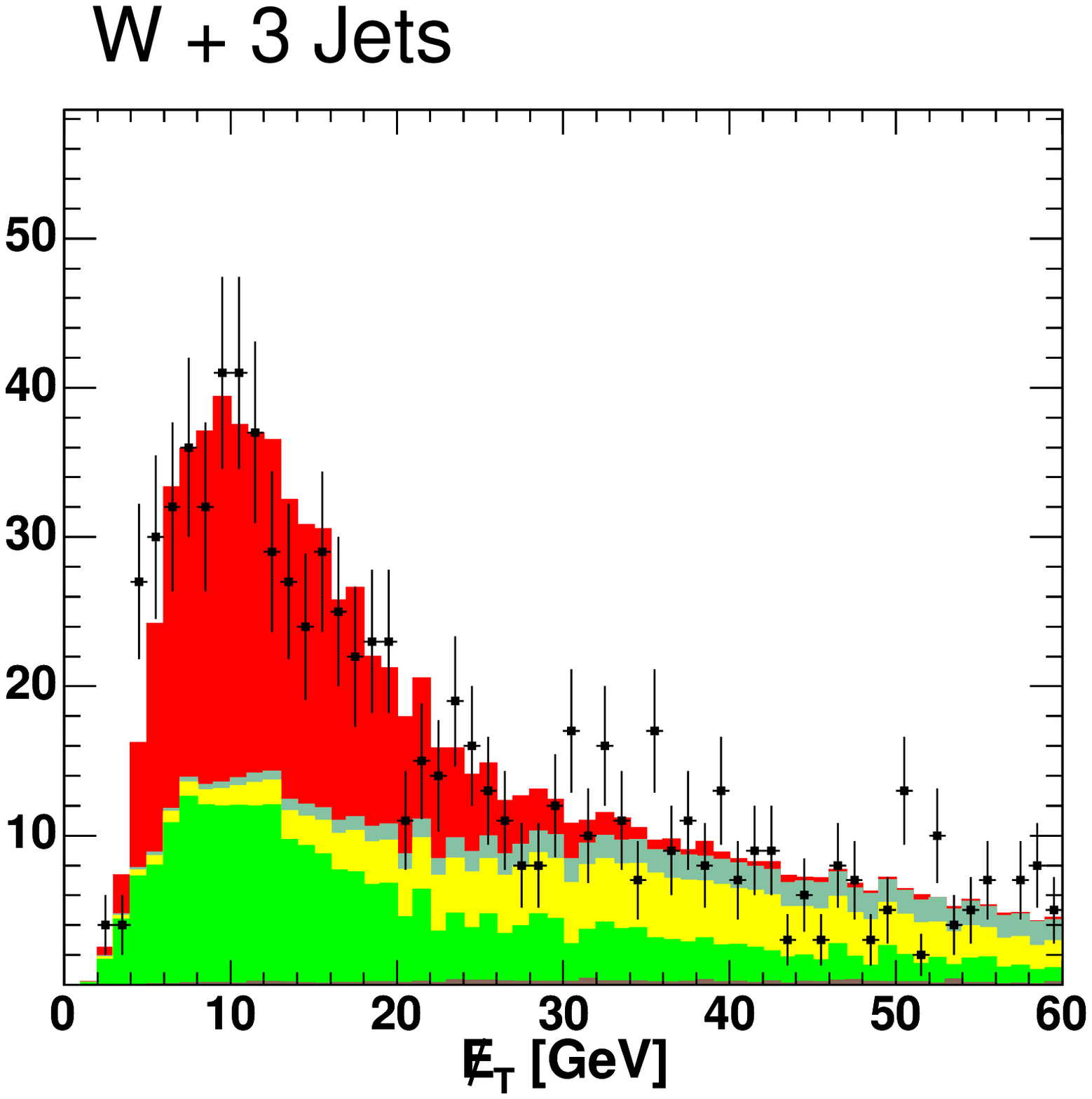}
\includegraphics[angle=0,width=0.45\textwidth]{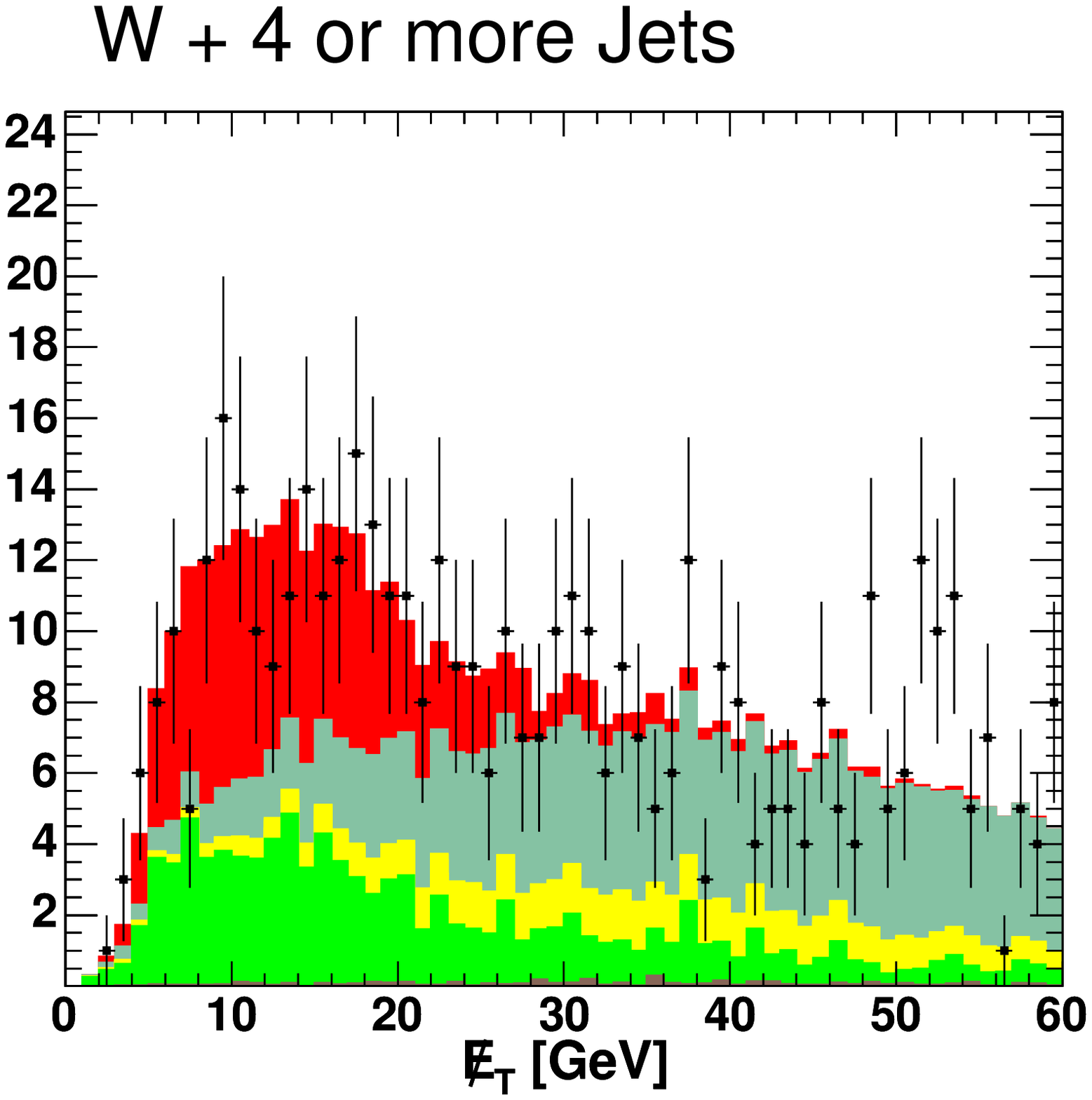}
\includegraphics[angle=0,width=0.45\textwidth]{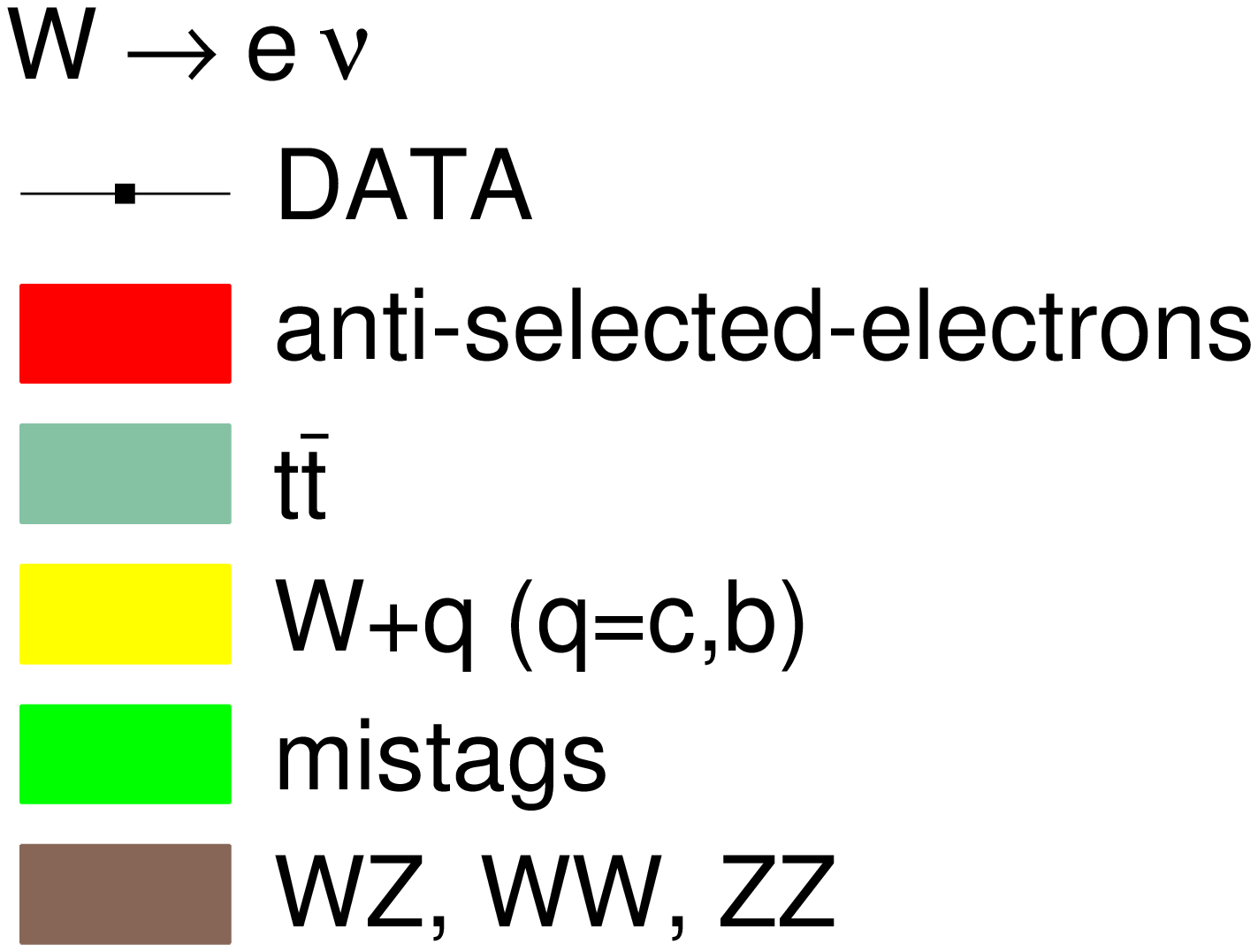}
\caption{The measured distribution in \met~for events with 1, 2, 3,
and 4 or more jets in events with a single tight electron and missing
energy forming a transverse mass, $M_{\mathrm{trans}}$, greater than
20 \GeV. At least one of the jets is required to be originating from a
heavy flavor quark. The observed distributions are compared to those
from SM expectations and non-$W$ events (labeled
``anti-selected-electron''- see text) in order to estimate the QCD
background. The order of stacking in the histograms is the same as in
the legend. }
\label{fig:fake_W}
\end{figure*}

 To estimate the non-$W$ background coming from jets we use
a data-derived model for non-$W$ events. The number of misidentified
$W$ bosons (non-$W$) is estimated separately for electrons and muons
by fitting the observed distributions in \met~with templates from real
$W$ decays and modeled non-$W$ events.  The distributions in \met~are
fitted over the range 0 $<$ \met~$<$ 60 GeV using events that contain
one tight lepton and no other leptons, with transverse mass
$M_{\mathrm{trans}}(\ell\nu)>20$ \GeV (see Fig.~\ref{fig:fake_W}).
For each jet multiplicity, the non-$W$ and the sum of the SM
contributions are separately normalized in the fit to the
\met~distribution.  The non-$W$ events are modeled by taking electrons
which pass all the selection criteria except those on the quality of
the calorimeter shower (labeled ``anti-selected-electrons'' in
Fig.~\ref{fig:fake_W}). The fractions of non-$W$ events are estimated
separately for events with 0, 1, 2, 3, and $\geq$4 jets in the final
state by propagating the distribution of the modeled non-$W$ events
into the region with \met~$>$ 25 GeV.  The estimated fractions of
non-$W$ events for each jet multiplicity are summarized in
Table~\ref{tab:Fake_W_Fract_incl}.
\begin{table}[h]
\centering
\caption{Fractions of non-$W$ events in events with one tight lepton
and no other leptons (inclusive $W$), and with \met~$>$25 GeV and
$M_{\mathrm{trans}}(\ell+\met)$ $>$ 20 GeV. Note that this sample is
selected without the requirement of the presence of a heavy-flavor
jet.}
\begin{tabular}{@{\extracolsep{\fill}}lccccc}
\hline
Jet Multiplicity & 0 jets &  1 jet &  2 jets &  3 jets  &  $\geq$4 jets \\
\hline
\hline
 \wenu~+ jets & 0.6\% & 1.9\% & 7\% &  14\%& 20\%\\
\hline
 \wmunu~+ jets & 0.1\% & 0.3\%  & 0.9\% & 1.8\%  & 2.6\% \\
\hline
\end{tabular}
\label{tab:Fake_W_Fract_incl}
\end{table}
\par A systematic uncertainty of 26\% is assigned on the fractions of
non-$W$ events~\citep{Cooper}, derived from the level of agreement
between the shape of the data-derived non-$W$ sample and the shape of
\met~distribution of misidentified electrons in data.

\subsection{Cosmic Ray Backgrounds}

\par High-energy cosmic muons traverse the CDF detector at a
significant rate and, if they intersect the beam-line, can be
reconstructed as $\mu^+\mu^-$ pairs. We remove cosmic ray events with
an algorithm which fits the two tracks of the $\mu^+\mu^-$ pair to a
single arc composed of an incoming track segment and an outgoing
segment, consistent in time evolution with a through-going
track~\citep{CRAY}. The algorithm also removes cosmic rays from events
where only one muon is reconstructed as a $W\goes \mu\met$ decay. It
searches for hits in the COT chamber within a narrow road along a
predicted trajectory opposite to the identified muon.  Finally, the
algorithm performs a simultaneous fit of the hits of the muon track
and the hits in the predicted trajectory with a single helix to
determine consistency with the cosmic-ray hypothesis.


\par
\begin{figure}[h]
\centering
\includegraphics[angle=0,width=0.45\textwidth]{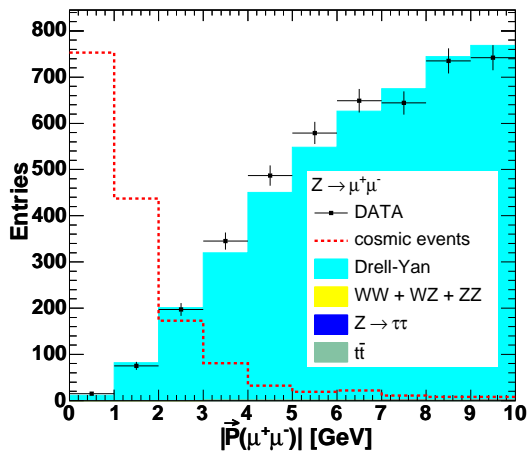}
\caption{The distribution $|\vec{P}(\mu^+\mu^-)|$ of muon pairs from
$Z$ boson candidates (solid), events expected from simulations
(stacked histogram), and from a cosmic ray sample (dashed). The
stacked histogram is mostly Drell-Yan production; the other sources
are considered but have negligible contribution to the histogram. The
number of cosmic ray events in the search sample is negligible.}
\label{fig:cosmic_momentum}
\end{figure}

 An independent estimate of the number of cosmic muons in the
$Z$ boson sample that have survived the cosmic-ray filter can be made
from the distribution of the magnitude of the momentum vector of the
$\mu^+\mu^-$ pair, $|\vec{P}(\mu^+\mu^-)|$. This is a simple way of
combining the usual ``back-to-back'' and momentum balance criteria for
the two muons into a single distribution, as cosmic $\mu^+\mu^-$ pairs
have a very narrow peak at $|\vec{P}(\mu^+\mu^-)|=0$ GeV, while real
\zmumu~decays occupy only a small area in the 3-dimensional momentum
phase space near $|\vec{P}(\mu^+\mu^-)|$=0. Using the
$|\vec{P}(\mu^+\mu^-)|$ distribution as an estimator, the number of
cosmic ray events in the sample surviving the cosmic filter is
negligible, as shown in Fig.~\ref{fig:cosmic_momentum}.

\section{Using $R$ as a Precise Check of the Monte Carlo Simulations}
\label{R_ratios}

\par The Monte Carlo simulation of a dataset extending over years,
with changing detector and accelerator conditions, is an exceptionally
complex task, involving large quantities of temporal conditions stored
in databases. Small errors in book-keeping or in properly specifying
which data to use in the code are difficult to detect. To validate the
modeling of the lepton identification, acceptances, and triggering we
use a calibration that is predicted to better than 1.5\% and is
directly sensitive to errors affecting overall efficiencies for
leptons and $\met$. We measure the ratio $R$ (see
Equation~\ref{R_equation}) of inclusively produced $W$ and $Z$ bosons
in their respective leptonic decay channels~\citep{R_ratio_exp}.

\par The ratio $R$ has been calculated at NNLO by F. A. Berends et
al., and is predicted to be 10.67 $\pm$ 0.15~\citep{Stirling_R}. We
measure $R$ = 10.52 $\pm$ 0.04(stat.) using electrons and 10.46 $\pm$
0.05(stat.) using muons. The observed numbers agree with the
theoretical prediction within 2\%, a negligible difference relative to
the other systematic uncertainties on the $t\goes Zc$ measurement.


\section{The FCNC Analysis}
\label{FCNC_Modeling}
\par Assuming that the FCNC decay of the top quark $t\goes Zc$ is
non-zero, a $t\bar{t}$ pair can decay to $WbWb$, $WbZc$, or $ZcZc$
with decay rates proportional to $(1-Br(t\goes Zc))^2$, $2Br(t\goes
Zc)\cdot(1-Br(t\goes Zc))$, and $Br(t\goes Zc)^2$, respectively. $W$
and $Z$ bosons are well identified via only their leptonic decay
modes, which have small branching fractions. To keep acceptances high
we require one of the bosons from the $t\bar{t}$ pair to decay
leptonically and the other hadronically.

\par To avoid large systematic uncertainties we analyze simultaneously
two final states from decays of top quark pairs: $p\bar{p}\goes
t\bar{t}\goes ZcWb\goes\ell\ell cjjb$ and $p\bar{p}\goes t\bar{t}\goes
WbWb\goes\ell\nu bjjb$, where: $\ell$ is a lepton ($e$ or $\mu$), $j$
is a jet, $\nu$ is a neutrino inferred via missing transverse energy
($\met$), $b$ and $c$ are ``heavy-flavor'' jets formed by
hadronization of a bottom-quark or a charm-quark, respectively. This
is done by comparing the number of expected events from SM $t\bar{t}$
decays and SM backgrounds to the number of observed events in each
final state. The contributions from $t\bar{t}$ decays depend on two
numbers: $Br(t\goes Zc)$ and $N_{t\bar{t}} = \sigma(p\bar{p}\goes
t\bar{t})\int{Ldt}$, where $\sigma(p\bar{p}\goes t\bar{t})$ is the
cross-section of top quark pair production at CDF and $\int{Ldt}$ is
the integrated luminosity.

\par Additional discrimination against SM backgrounds is achieved by
requiring at least one of the four jets in the final state to be
consistent with originating from a heavy-flavor quark ($b$ or $c$
quark). The identification of a heavy-flavor jet is performed with the
``$b$-tagging'' algorithm which is introduced earlier in
Section~\ref{btag}.
\par The unknown structure of the FCNC coupling is parameterized via
the polarization of the $Z$ boson produced in $t\goes Zc$ decay, as
the polarization is the only parameter that affects the acceptance of
FCNC top quark decays. We vary the value of the longitudinal
polarization of the $Z$ bosons from 0.0 to 1.0. The final result is
presented as a function of the longitudinal polarization.
\par We reconstruct the invariant mass of the top quark,
$M_{\mathrm{top}}$, in events with two leptons and four jets assuming
that the events are $t\bar{t}$ FCNC decays.  The distribution of
$M_{\mathrm{top}}$ provides additional separation between standard
model backgrounds and the FCNC signal; the top quark mass,
$M_{\mathrm{top}}$, distribution for background events peaks below the
FCNC signal.

\section{Measuring Top Quark Pair Production in Events with a $W$
Boson and Four Jets}
\label{ttbar}

\par The measurement of the FCNC branching ratio relies on two
datasets (see Section~\ref{analysis}): $\ell\ell + 4~\mathrm{jets}$
and $\ell\mmet + 4~\mathrm{jets}$, where $\ell\ell$ and $\ell\mmet$
are consistent with decays of a $Z$ boson or a $W$ boson (see
Sections~\ref{zdef} and~\ref{wdef}), respectively. In this section we
focus only on events with $\ell\mmet + 4~\mathrm{jets}$, where the
majority comes from $t\bar{t}\goes WbWb$ decays. At least one of the
four jets in the final state is required to be identified as heavy
flavor (HF) decay by the secondary vertex identification
algorithm. The estimate of SM production of $W$+HF events
(e.g. $W+b\bar{b}$) requires normalization of three key components:
\ttbar; $W+b\bar{b}$, $W+c\bar{c}$, $W+c$; and ``non-$W$'' background
events, which arise from mis-measured jet events.

\subsection{Estimating the Contributions from \ttbar~Production, $W$+HF
Production, and from non-$W$ Backgrounds}


\par The dominant SM contribution to the $W$+4-jet bin with one jet
identified as heavy flavor (a ``$b$-tag'') is \ttbar~production. The
production of a $W$ boson with heavy flavor, $W+b\bar{b}$,
$W+c\bar{c}$, and $W+c$, however, dominates production in the $W$+2
jet bin. We consequently use the spectrum in the number of jets in
$W$+HF production to estimate the contribution from \ttbar~alone in an
iterative process. We take the top quark pair production cross-section
to be $\sigma(t\bar{t})=~7.6$ pb~\citep{top_76pb}.

\par We initially assume that the fraction of non-$W$ events is
negligible. We determine the normalization of the standard model
contribution to the $W$+HF processes $W$+$b\bar{b}$, $W+c$
and $W+c\bar{c}$ by rescaling the respective cross-sections to
match the total number of events observed in the $W$ + 2 jets
bin. We assume that the overall normalization of $W+b\bar{b}$ +
jets, $W+c\bar{c}$ + jets, and $W+c$ + jets can be corrected by
a single scale factor that is the same for the electron and muon
channels.
\par We then use this normalization of the $W$+HF samples to
estimate the remaining contribution from non-$W$'s, as described in
detail below in Section~\ref{fake_W_HF}.
\par
\begin{figure}[h]
\centering
\includegraphics[angle=0,width=0.45\textwidth]{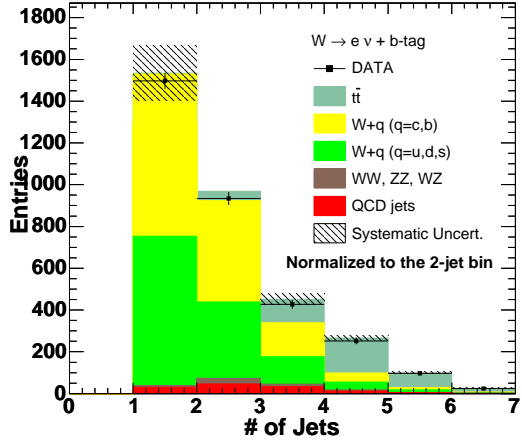}
\includegraphics[angle=0,width=0.45\textwidth]{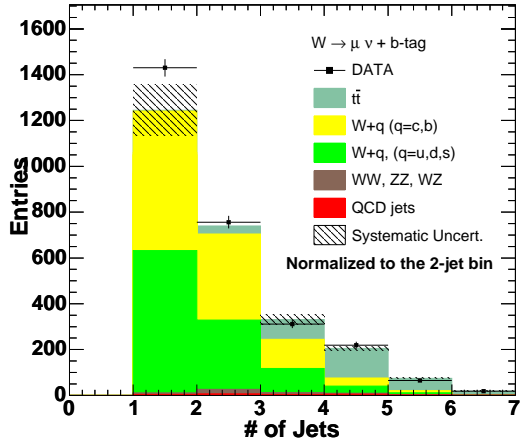}
\caption{The measured distributions (points) in the number of jets in
events with a $W$ and a $b$-tag for \wenu~and \wmunu, compared to SM
expectations (histogram). The order of stacking in the histograms is
the same as in their legends. }
\label{fig:W_BTG_NJets}
\end{figure}

 We then repeat the calculation of the fraction of real $W$+HF
events using the estimate of non-$W$'s, rescaling of the $W$+HF by a
factor of $0.97~\pm~0.09$ to match the number of events in the $W$ + 2
jets bin.  The final jet multiplicity distributions for the $W$ + HF
sample are shown in Fig.~\ref{fig:W_BTG_NJets}. We find good agreement
for events with three or more jets in the $W$+HF sample.
\par The motivation for normalizing to the two-jet multiplicity bin is
based on the matrix-element structure of associated heavy flavor
production in $W$ and $Z$ events. A problem with any normalization
scheme that uses the 1-jet bin is that different diagrams contribute
to the N$=$1 and the N$=$2 jet multiplicity bins; taking into account
the (large, particularly for charm) NLO corrections is tricky since
the corrections differ significantly for the different processes. In
contrast, the radiation of additional jets and jet matching procedures
in the higher-multiplicity jet bins are fairly well
understood~\citep{CKKW} in comparison to the uncertainties in the
1-jet bin. We avoid these issues by normalizing the multi-jet
multiplicity distribution to the 2-jet bin.

\par
\begin{figure}[h]
\centering
\includegraphics[angle=0,width=0.45\textwidth]{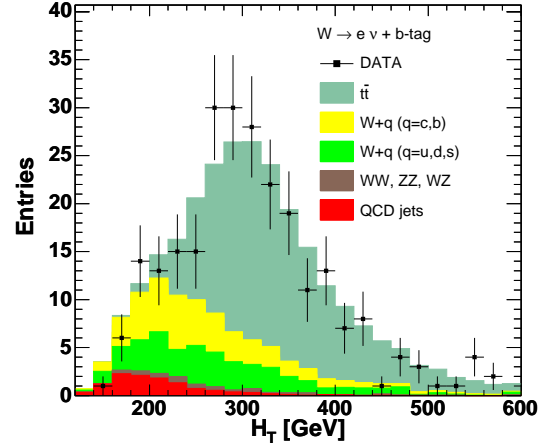}
\includegraphics[angle=0,width=0.45\textwidth]{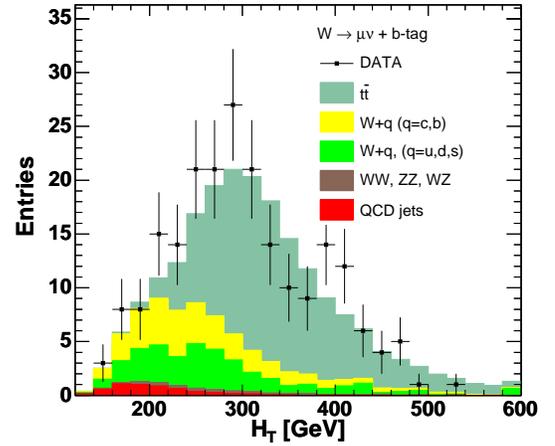}
\caption{The measured distribution (points) in $H_T$ in events with a
W and a $b$-tag, compared to SM expectations (histogram), for the
electron channel (upper figure) and muon channel (lower figure).  The
order of stacking in the histograms is the same as in their legends.}
\label{fig:W_BTG_Ht_4J}
\end{figure}

 We perform an additional consistency check by comparing a
measured top quark pair production cross-section with its theoretical
prediction, assuming that there are no FCNC~\citep{no_signal}. In the
$W$ + 4-jet bin the ratio of the measured cross-section to the SM
expectation is 1.17 $\pm$ 0.09, where the SM background is evaluated
using a top quark cross-section of 7.6 pb and $Br(t\goes Wb)$ = 100\%
(i.e $Br(t\goes Zc)$ = 0). The $H_T$-distribution for the $W$ + 4 jets
events agree well with those of top quark pair decays (see
Fig.~\ref{fig:W_BTG_Ht_4J}), where the contribution from the top quark
pair production is normalized with the measured cross-section.  The
total transverse energy, $H_T$, is a scalar sum of $E_T$ of all
reconstructed objects (electrons, muons, photons, jets, missing
transverse energy, and unclustered energy). The transverse energies of
the objects are the same as those used for the calculation of the
missing energy $\met$ in the event.

\subsection{Non-$W$ Backgrounds in the \ttbar~Sample}
\label{fake_W_HF}
\par The same procedure used for measuring background in the inclusive
$W$ bosons sample (see Section~\ref{w_qcd}) is used to measure
backgrounds in the \ttbar~sample. The number of misidentified $W$
bosons (non-$W$'s) is estimated by fitting the \met-distribution for
each jet multiplicity bin in events with one tight lepton and
$M_{\mathrm{trans}}$ higher than 20 GeV, where the transverse mass
$M_{\mathrm{trans}}$ is calculated for the lepton and \met.  The
fractions of non-$W$ events obtained after applying the \met-cut
(\met$>$25 GeV) are presented in Table~\ref{tab:Fake_W_Fract_BTG}
versus the jet multiplicity~\citep{four_jet_fit}.
\begin{table}[h]
\centering
\caption{The fractions of non-$W$ QCD background (labeled as QCD-jets
in Fig.~\ref{fig:W_BTG_NJets}) in events with a tight lepton ($e$ or
$\mu$), \met~$>$25 GeV, $M_{\mathrm{trans}}(\ell+\met)$ $>$ 20 GeV,
and at least one $b$-tagged jet.}
\begin{tabular}{lcccc}
\hline
Jet Multiplicity & 1 jet &  2 jets &  3 jets  &  $\geq$4 jets \\
\hline
\hline
\wenu~+ jets &  2.0\% & 4.9\% & 7.6\%& 4.7\%\\
\hline
\wmunu~+ jets & 0.3\%  & 0.9\% & 1.3\%  & 2.6\% \\
\hline
\end{tabular}
\label{tab:Fake_W_Fract_BTG}
\end{table}
\par The acceptance times efficiency, $A_{WW\goes\ell\mmet}$ (see
Subsection~\ref{FCNC_accept}), is determined from the MC simulations
of the standard model \ttbar~decays for the $W$ + 4 jets bin. The
obtained numbers are presented in Table \ref{tab:w4j_efficiency}. The
cumulative acceptances $A_{WW\goes\ell\mmet}$ include the branching
fractions for $W\goes\ell\nu$ and $W\goes qq'$ decays.
\begin{table}[h]
\centering
\caption{The acceptance times efficiency for $\ell\mmet$ + 4 jets
events which are produced via $t\bar{t}\goes WbWb\goes\ell\mmet$ +
4 jets decay chain (see Subsection~\ref{FCNC_accept}). The
efficiencies are calculated from the Monte Carlo simulations and
corrected to match the lepton identification and triggering
efficiencies in data.}
\begin{tabular}{lc}
\hline
Process & $A_{WW\goes\ell\mmet}$  \\
\hline
\hline
\ttbar $\goes WbWb \goes e\met +$ 4 jets & 0.0128 \\
\hline
\ttbar $\goes WbWb \goes \mu\met +$ 4 jets  & 0.00994  \\
\hline
\end{tabular}
\label{tab:w4j_efficiency}
\end{table}

\subsection{Summary of the Backgrounds in W+4 Jets}
\label{summary_W_4jets}
\par The number of events observed and the expected number from all
processes except \ttbar~production are given in
Table~\ref{tab:w4j_observed}.
\begin{table}[h]
\centering
\caption{A summary of the numbers of $W$ + 4 jets events. At least
one jet in each event is required to be $b$-tagged. }
\begin{tabular}{lp{0.75in}c}
\hline
Final state & \centering Observed & Background (non-\ttbar) \\
\hline
\hline
$e\met+ 4~\mathrm{jets}$ &\centering 252 & 98.7  \\
\hline
$\mu\met+4~\mathrm{jets}$ &\centering 219 & 75.2  \\
\hline
\end{tabular}
\label{tab:w4j_observed}
\end{table}

\section{The Contribution from FCNC Decays of \ttbar~Pairs to Events 
with $W$/$Z$ Bosons and Jets}
\label{FCNC_Wjets}
\subsection{Acceptances for \ttbar~Decays}
\label{FCNC_accept}
\par We use a modified version of the {\sc madgraph} Monte Carlo event
generator \citep{MadGr} to produce simulated events for the
$t\bar{t}\goes ZcWb$ and $t\bar{t}\goes ZcZc$ processes, which are
then hadronized using {\sc pythia}.
%
%
\par In order to calculate the rates of expected events for the two
final states $\ell\ell$ + 4 jets and $\ell\met$ + 4 jets,
we need to introduce a notation for the acceptances multiplied by
efficiencies, $(A\cdot\epsilon)_{\mathrm{Y}}$, for the decay chain ``Y''
of \ttbar~pairs. Acceptance $(A\cdot\epsilon)_{\mathrm{Y}}$ is a fraction
of \ttbar~events observed in the corresponding final state. The
acceptances $(A\cdot\epsilon)_{\mathrm{Y}}$ include combinatoric factors
and the corresponding branching fractions for decays of $W$'s and
$Z$'s: $Br(W\goes\ell\nu)$, $Br(W\goes qq')$, $Br(Z\goes\ell\ell)$,
and $Br(Z\goes q\bar{q})$. 

\par The acceptances $(A\cdot\epsilon)_{\mathrm{Y}}$ depend on the FCNC
branching ratio $Br(t\goes Zc)$. We divide the acceptances by
polynomials dependent on $Br(t\goes Zc)$ to factor out the terms
independent of the FCNC branching ratio:
\begin{eqnarray}
A_{ZZ\goes\ell\ell} =&&\frac{(A\cdot\epsilon)_{t\bar{t}\goes ZcZc\goes
    \ell\ell+4~\mathrm{jets}}}{Br(t\goes Zc)^2}, \\
A_{ZW\goes\ell\ell} =&&\frac{(A\cdot\epsilon)_{t\bar{t}\goes ZcWb\goes\ell\ell+4~\mathrm{jets}}}
{Br(t\goes Zc)\cdot(1-Br(t\goes Zc))}, \\
A_{WZ\goes\ell\mmet} =&& \frac{(A\cdot\epsilon)_{t\bar{t}\goes ZcWb\goes\ell\nu+4~\mathrm{jets}}}
{Br(t\goes Zc)\cdot(1-Br(t\goes Zc))}+\nonumber\\
                && \frac{(A\cdot\epsilon)_{t\bar{t}\goes ZcWb\goes\ell\mmet+4~\mathrm{jets}}}
{Br(t\goes Zc)\cdot(1-Br(t\goes Zc))}, \\
A_{WW\goes\ell\mmet} =&&\frac{(A\cdot\epsilon)_{t\bar{t}\goes WbWb\goes \ell\nu+4~\mathrm{jets}}}{(1-Br(t\goes Zc))^2},
\end{eqnarray}
and
\begin{equation}
A_{ZZ\goes\ell\mmet} = \frac{(A\cdot\epsilon)_{t\bar{t}\goes ZcZc\goes\ell\mmet+4~\mathrm{jets}}}{Br(t\goes Zc)^2}.
\end{equation}
The values of $A_{\mathrm{Y}}$ are determined using simulated samples
where all the $t\bar{t}$ pairs decay exclusively to only one of the
intermediate states: $WbZc$, $ZcZc$, or $WbWb$. The acceptance
$A_{WZ\goes\ell\mmet}$ includes two decay chains since the missing
energy, \met, can be produced via decay $W\goes\ell\nu$ or by
mis-identifying $Z\goes\ell\ell$ decay. The $Z\goes\ell\ell$ decay can
be mistaken for a $W\goes\ell\nu$ decay when one of the two leptons is
not identified (i.e. missing). The loss of the real lepton can create
significant missing energy.

\subsection{Properties of the FCNC $t\goes Zc$ Coupling}
\par We note that the helicity structure of a possible $t\goes Zc$
vertex is model-dependent. We cover the full range of possible
helicities so as to be assumption-independent.
\par The kinematic properties of $t\goes Zc$ decay are reflected by
the angular distributions of the decay products. This affects the
total acceptance for the FCNC events since the isolation requirement
is placed on all the identified jets and leptons. For example, the
final state of the $t\goes Zc\goes\ell\ell c$ decay chain can be fully
described by introducing an angle $\theta^*$, taken to be the angle
between the direction of the top quark (anti-top-quark) and the
positive (negative) lepton in the rest frame of the $Z$ boson. The
angular distribution of $\theta^*$ has the following general form:
\begin{equation}
f(\theta^*) = a_0\cdot f_0(\theta^*) + a_1\cdot f_1(\theta^*) + a_2\cdot f_2(\theta^*),
\end{equation}
where $a_0$, $a_1$, and $a_2$ are constants which depend on the
polarization of the $Z$ boson and whose sum is one
($a_0+a_1+a_2=1$). The functions $f_i(\theta^*)$ are given by:
\begin{equation}
f_0(\theta^*) = \frac{3}{4}(1-\cos^2(\theta^*)),
\end{equation}
\begin{equation}
f_1(\theta^*) = \frac{3}{8}(1+\cos(\theta^*))^2,
\end{equation}
and
\begin{equation}
f_2(\theta^*) = \frac{3}{8}(1-\cos(\theta^*))^2.
\end{equation}

\par The angular distribution of decay products of the $t\goes
Wb\goes\ell\nu b$ decay is parametrized with the same function
$f(\theta^*)$ by taking appropriate values of the $a_i$. In the case
of $t\goes Wb$ decay the coefficients $a_0$, $a_1$, and $a_2$ are the
fractions of longitudinal, left-handed, and right-handed helicities of
the $W$ boson, respectively. However, the $Z$ boson, unlike the
$W$ boson, has both right-handed and left-handed couplings.
Consequently, while the coefficient $a_0$ is simply the fraction of
the longitudinally polarized $Z$ bosons, the coefficients $a_1$ and
$a_2$ are linear functions of the fractions of left-handed and
right-handed helicities of the $Z$ boson.

\par The distribution of $\cos(\theta^*)$ resulting from an arbitrary
FCNC coupling can always be described by choosing appropriate values
for the constants $a_i$.  The acceptances of the FCNC top quark decays
$A_{\mathrm{Y}}$ depend on the angular distributions of the decay
products since we require the isolation in a cone of 0.4 for all the
identified leptons and jets. In consequence, the acceptances are
functions of $a_0$ and $a_1$ (i.e. $A_{\mathrm{Y}} =
A_{\mathrm{Y}}(a_0,a_1$), noting that $a_2=1-a_0-a_1$). The top quark
decay is symmetric with respect to the charge of the fermion
($\ell\bar{\ell}$ or $q\bar{q}$), and therefore the acceptances
calculated for decays of right-handed $Z$ bosons and left-handed
bosons are identical. This means that the acceptances $A_{\mathrm{Y}}$
can be fully parametrized with the fraction of longitudinally
polarized $Z$ bosons (i.e.  $A_{\mathrm{Y}} =
A_{\mathrm{Y}}(a_0,1-a_0) = A_{\mathrm{Y}}(a_0)$).

\par We compute each acceptance $A_{\mathrm{Y}}$ for five values of
the fraction of longitudinally polarized $Z$ bosons using Monte Carlo
simulated events. This allows us to calculate the acceptances
$A_{\mathrm{Y}}$ for any fraction $a_0$ by interpolating the
acceptances $A_{\mathrm{Y}}$ between the points measured. The
acceptance $A_{WW\goes\ell\mmet}$ is a constant since it does not
have any FCNC vertices. The other acceptances,
$A_{ZZ\goes\ell\ell}$, $A_{ZW\goes\ell\ell}$,
$A_{WZ\goes\ell\mmet}$, and $A_{ZZ\goes\ell\mmet}$ have linear or
quadratic dependences on the fraction of the longitudinal helicity of
the $Z$ bosons:
\begin{widetext}
\begin{equation}
A_{ZZ\goes\ell\ell}(a_0) = a_0^2 \cdot A_{ZZ\goes \ell\ell}^{\mathrm{long}} +2\cdot a_0\cdot(1-a_0)\cdot
A_{ZZ\goes\ell\ell}^{\mathrm{corr}} + (1-a_0)^2\cdot A_{ZZ\goes \ell\ell}^{\mathrm{left}},
\end{equation}
\begin{equation}
A_{ZW\goes\ell\ell}(a_0) = a_0\cdot A_{ZW\goes\ell\ell}^{\mathrm{long}}+(1-a_0)\cdot A_{ZW\goes \ell\ell}^{\mathrm{left}},
\end{equation}
\begin{equation}
A_{WZ\goes\ell\mmet}(a_0) = a_0\cdot A_{WZ\goes\ell\mmet}^{\mathrm{long}}+(1-a_0)\cdot A_{WZ\goes\ell\mmet}^{\mathrm{left}},
\end{equation}
and
\begin{equation}
A_{ZZ\goes\ell\mmet}(a_0) = a_0^2 \cdot A_{ZZ\goes\ell\mmet}^{\mathrm{long}}+2\cdot a_0\cdot(1-a_0)\cdot A_{ZZ\goes\ell\mmet}^{\mathrm{corr}}
 + (1-a_0)^2\cdot A_{ZZ\goes\ell\mmet}^{\mathrm{left}},
\end{equation}
\end{widetext}
where $A_{\mathrm{Y}}^{\mathrm{long}}$ are measured for the
longitudinally-polarized component of the $Z$ decays,
$A_{\mathrm{Y}}^{\mathrm{left}}$ are for the left-handed component,
and the value of $A_{\mathrm{Y}}^{\mathrm{corr}}$ is obtained using
FCNC events where the $Z$ bosons are mixed with 50\% left-handed and
50\% longitudinal polarizations. The acceptance
$A_{ZZ\goes\ell\ell}$ has a quadratic dependence on $a_0$ since it
accounts for the two FCNC decays of the top and anti-top quarks. The
numerical values of the acceptances are tabulated in
Sections~\ref{ttbar} and~\ref{FCNC_Zjets}.

\section{Measuring the Contributions from FCNC and SM 
Processes in Events with a $Z$ Boson and Four Jets}

\label{FCNC_Zjets}
\par
\begin{figure}[h]
\centering
\includegraphics[angle=0,width=0.45\textwidth]{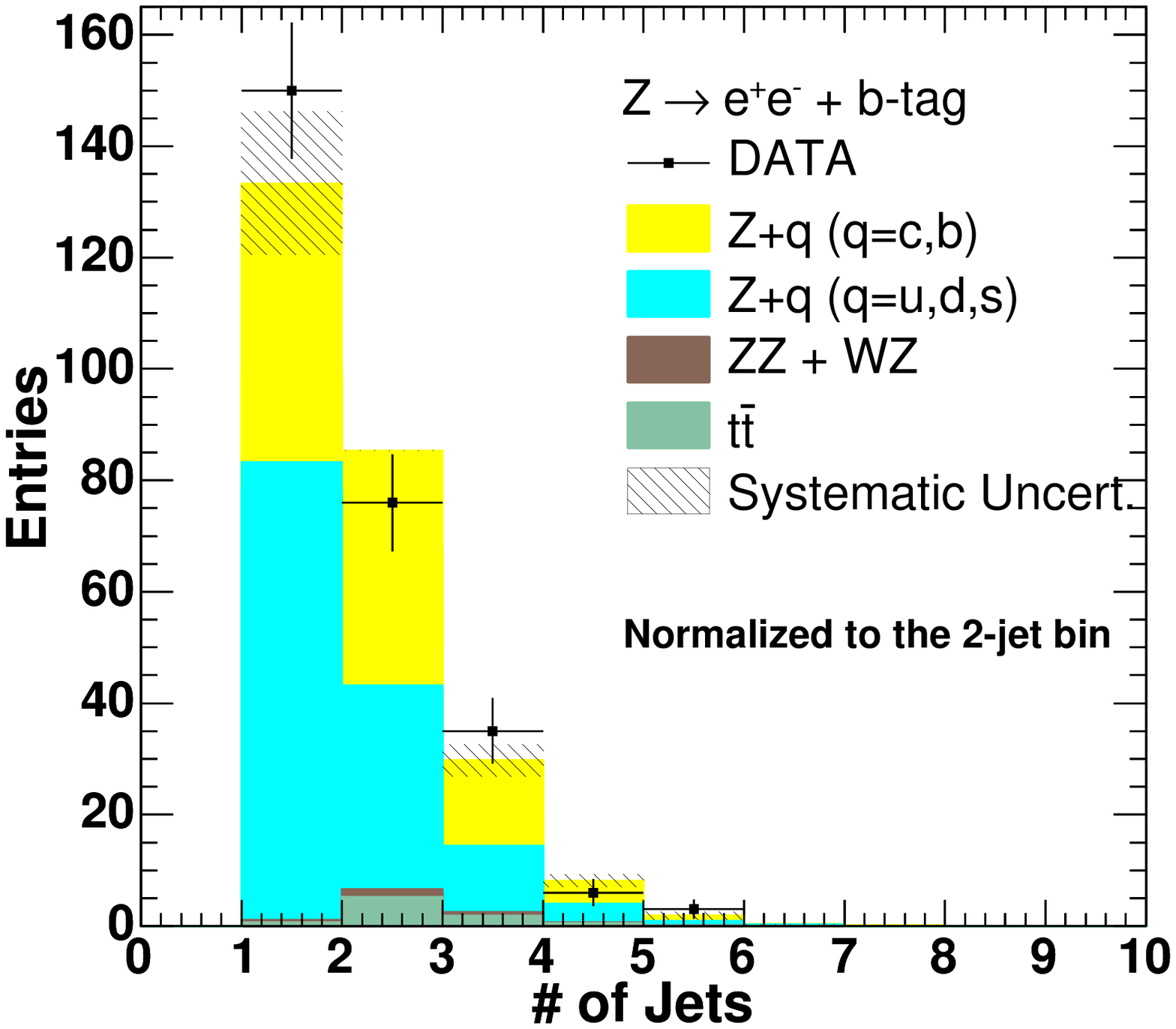}
\includegraphics[angle=0,width=0.45\textwidth]{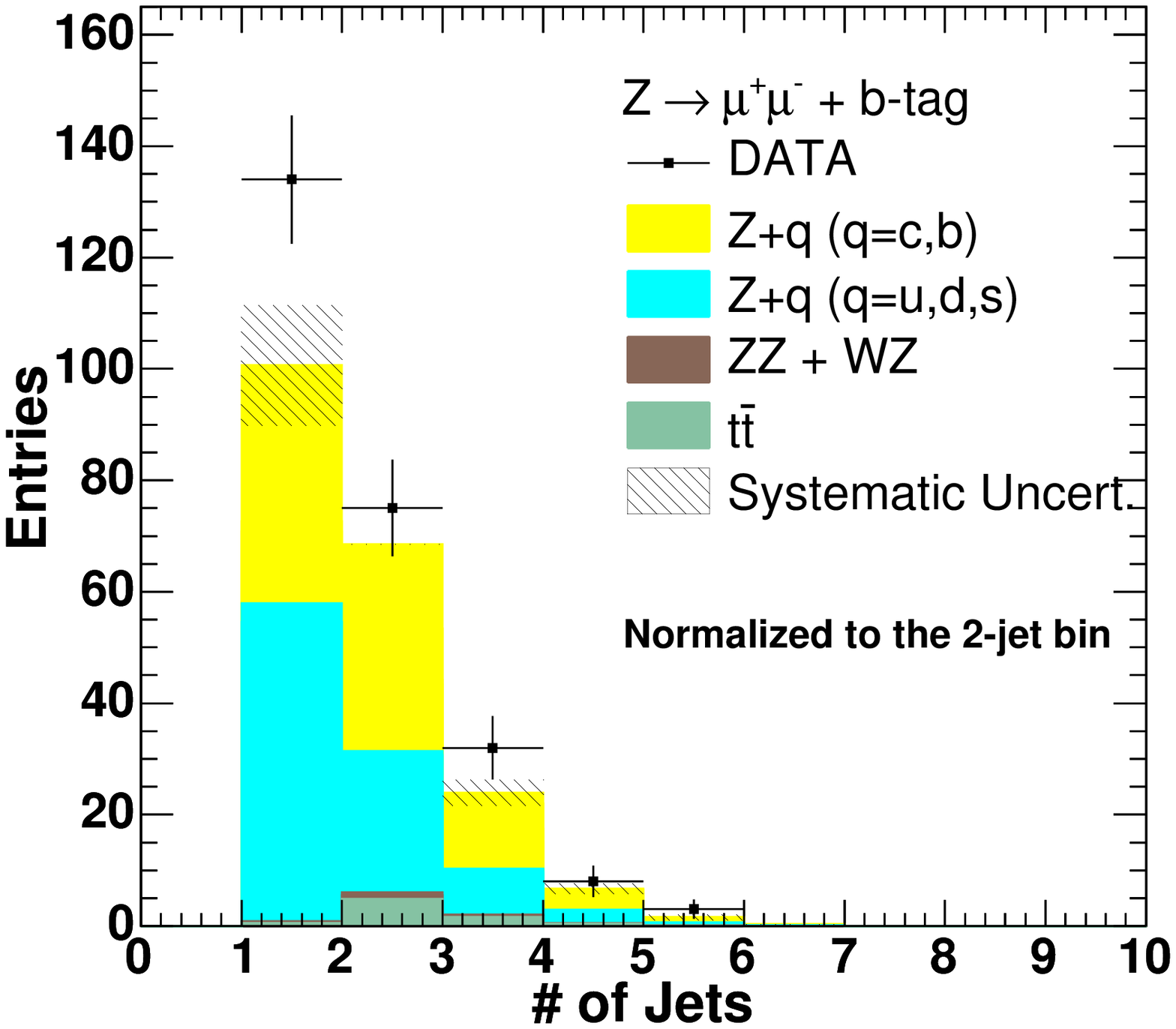}
\caption{The measured distribution (points) in the number of jets in
events with a $Z$ and a $b$-tag, compared to SM expectations
(histogram), for the electron channel (upper figure) and muon channel
(lower figure). We normalize to the average of the \zee~and \zmumu~2
jet bins. The order of stacking in the histograms is the same as in
their legends. }
\label{fig:Z_BTG_NJets}
\end{figure}

 At this stage we consider only events which have two leptons
consistent with a parent $Z$ boson and at least one $b$-tagged jet. We
use the jet multiplicity distribution (see Fig.~\ref{fig:Z_BTG_NJets})
to constrain the number of non-SM $Z$+4-jet events. We do this by
scaling the total $Z$+HF component, $Z$+$q$ ($q$=$c,b$), to the number
of (observed - mis-tagged) $Z$+2 jets events in the electron and muon
modes simultaneously. The fraction of mis-tagged, $Z$+$q$ ($q=u,d,s$),
events is estimated from data using inclusive $Z$ + jets events (see
Section~\ref{btag}).
The number of $Z$ + 4 jets events observed and the expected number
from all SM processes are given in Table~\ref{tab:z4j_observed}.
\begin{table}[h]
\centering
\caption{A summary of the numbers of $Z$ + 4 jets events. At least
one jet in each event is required to be $b$-tagged. }
\begin{tabular}{lcc}
\hline
Final state & Observed & SM Background \\
\hline
\hline
$e^+e^-+ 4~\mathrm{jets}$ & 6 & 8.4  \\
\hline
$\mu^+\mu^-+4~\mathrm{jets}$ & 8 & 6.9  \\
\hline
\end{tabular}
\label{tab:z4j_observed}
\end{table}


\par The FCNC signal contribution is divided into two parts, $ZcWb$
and $ZcZc$, since the $b$-tagging rates are different. We summarize
the acceptance and the efficiency measurements for the $Z$ + 4 jets
channel in Tables~\ref{tab:zczc4j_efficiency},
\ref{tab:zcwb4j_efficiency}, \ref{tab:wbzc4j_efficiency_a3}, and
\ref{tab:zczc4j_efficiency_a5}. The case when a leptonic decay of a
$Z$ boson is misidentified as the leptonic decay of a $W$ boson is taken 
into account in Tables \ref{tab:wbzc4j_efficiency_a3} and
\ref{tab:zczc4j_efficiency_a5}.

\par The top quark mass $M_{\mathrm{top}}$ is used as a discriminating
variable against the SM backgrounds as was mentioned earlier in
Section~\ref{FCNC_Modeling}. More details on the calculation of
$M_{\mathrm{top}}$ are provided later in Section~\ref{fitting_top}.
We recalculate the acceptances $A_{ZZ\goes\ell\ell}^{i}$ and
$A_{ZW\goes\ell\ell}^{i}$ for the $i$'th bin of the top quark mass
distribution by multiplying a cumulative acceptance $A_{\mathrm{Y}}$
(Y is $ZZ\goes\ell\ell$ or $ZW\goes\ell\ell$) and the fraction of
events in the $i$'th bin:
\begin{equation}
A_{\mathrm{Y}}^{i} = A_{\mathrm{Y}}\cdot\frac{N_i}{\displaystyle\sum_{k} N_k}.
\end{equation}
The obtained acceptances $A^i_{\mathrm{Y}}$ depend on the reconstructed top
quark mass of the \ttbar~FCNC events.
\begin{table}[h]
\centering
\caption{The acceptance times efficiency for the dilepton signature
from the inclusive FCNC decay of \ttbar $\goes
ZcZc\goes\ell\ell+ccjj$ for different values of the longitudinal
fraction of $Z$ bosons, for electron pairs and muon pairs
separately. The SM branching ratios for the $Z\goes \ell\ell$
decays are included.
}
\begin{tabular}{lc}
\hline
Process & $A_{ZZ\goes\ell\ell}$\\
\hline
\hline
\multicolumn{2}{c}{ Longitudinal fraction is $a_0$=0.00 }\\
\hline
\ttbar $\goes ZcZc \goes e^+e^-+4~\mathrm{jets}$ & 0.00185 \\
\hline
\ttbar $\goes ZcZc \goes \mu^+\mu^-+4~\mathrm{jets}$ & 0.00178 \\
\hline
\hline
\multicolumn{2}{c}{ Longitudinal fraction is $a_0$=0.50 }\\
\hline
\ttbar $\goes ZcZc \goes e^+e^-+4~\mathrm{jets}$ & 0.00203 \\
\hline
\ttbar $\goes ZcZc \goes \mu^+\mu^-+4~\mathrm{jets}$ & 0.00192 \\
\hline
\hline
\multicolumn{2}{c}{ Longitudinal fraction is $a_0$=1.00}\\
\hline
\ttbar $\goes ZcZc \goes e^+e^-+4~\mathrm{jets}$ & 0.00222 \\
\hline
\ttbar $\goes ZcZc \goes \mu^+\mu^-+4~\mathrm{jets}$ & 0.00205 \\
\hline
\hline
\end{tabular}
\label{tab:zczc4j_efficiency}
\end{table}
\begin{table}[h]
\centering
\caption{A summary of the acceptance times efficiency for the dilepton
signature from inclusive FCNC decays of \ttbar $\goes ZcWb\goes
\ell\ell + bcjj$, for different values of the longitudinal fraction 
of $Z$ bosons, for electron pairs and muon pairs separately.
}
\begin{tabular}{lc}
\hline
Process & $A_{ZW\goes\ell\ell}$ \\
\hline
\hline
\multicolumn{2}{c}{ Longitudinal fraction is $a_0$=0.00 }\\
\hline
\ttbar $\goes ZcWb\goes e^+e^-+4~\mathrm{jets}$ & 0.00275  \\
\hline
\ttbar $\goes ZcWb\goes\mu^+\mu^-+4~\mathrm{jets}$ & 0.00267 \\
\hline
\hline
\multicolumn{2}{c}{ Longitudinal fraction is $a_0$=1.00 }\\
\hline
\ttbar $\goes ZcWb \goes e^+e^-+4~\mathrm{jets}$ & 0.00313 \\
\hline
\ttbar $\goes ZcWb \goes \mu^+\mu^-+4~\mathrm{jets}$ & 0.00293 \\
\hline
\hline
\end{tabular}
\label{tab:zcwb4j_efficiency}
\end{table}
\begin{table}[h]
\centering
\caption{A summary of the acceptance times efficiency for the
contribution to the single lepton+$\met$ signature from the inclusive
FCNC decays of \ttbar $\goes WbZc \goes \ell\met+bcjj$ (i.e. the decay
of a $Z$ boson is mis-identified as the decay of a $W$ boson) and
\ttbar $\goes WbZc \goes \ell\nu+bcjj$. Standard model branching
ratios are included. The acceptance $A_{WZ\goes\ell\mmet}$ is the sum
of acceptances for the decay modes which contribute to the
signature of $\ell + \met + 4~\mathrm{jets}$.}
\begin{tabular}{lc}
\hline
Process & $A_{WZ\goes\ell\mmet}$ \\
\hline
\hline
\multicolumn{2}{c}{ Longitudinal fraction is $a_0$=0.00 }\\
\hline
\ttbar $\goes WbZc \goes e\nu$ + 4 jets & 0.00927\\
\hline
\ttbar $\goes WbZc \goes e\met$ + 4 jets & 0.00179\\
\hline
\ttbar $\goes WbZc \goes \mu\nu$ + 4 jets & 0.007915 \\
\hline
\ttbar $\goes WbZc \goes \mu\met$ + 4 jets & 0.002180 \\
\hline
\hline
\multicolumn{2}{c}{ Longitudinal fraction is $a_0$=1.00 }\\
\hline
\ttbar $\goes WbZc \goes e\nu$ + 4 jets   & 0.00967 \\
\hline
\ttbar $\goes WbZc \goes e\met$ +4 jets  & 0.00185 \\
\hline
\ttbar $\goes WbZc \goes \mu\nu$ + 4 jets  & 0.00817 \\
\hline
\ttbar $\goes WbZc \goes \mu\met$ + 4 jets & 0.00227 \\
\hline
\hline
\end{tabular}
\label{tab:wbzc4j_efficiency_a3}
\end{table}
\begin{table}[th]
\centering
\caption{The acceptance times efficiency for the contribution to the
single lepton+$\met$ signature from the inclusive FCNC decay of \ttbar
$\goes ZcZc \goes \ell+\mmet+ccjj$, where at least one di-leptonic
decay of $Z$ boson has been mis-identified as the decay of a $W$
boson. Standard model branching ratios are included. Note that this
channel depends on the square of the FCNC branching ratio for the $Z$,
and so its contribution is suppressed relative to that from the case
where only one $Z$ decays by FCNC (see
Table~\ref{tab:wbzc4j_efficiency_a3}).}
\begin{tabular}{lc}
\hline
Process & $A_{ZZ\goes\ell\mmet}$\\
\hline
\hline
\multicolumn{2}{c}{ Longitudinal fraction is $a_0$=0.00 }\\
\hline
\ttbar $\goes ZcZc \goes e+\mmet+4~\mathrm{jets}$ & 0.000873  \\
\hline
\ttbar $\goes ZcZc \goes \mu+\mmet+4~\mathrm{jets}$ & 0.00127 \\
\hline
\hline
\multicolumn{2}{c}{ Longitudinal fraction is $a_0$=0.50 }\\
\hline
\ttbar $\goes ZcZc \goes e+\mmet+4~\mathrm{jets}$ & 0.000858 \\
\hline
\ttbar $\goes ZcZc \goes \mu+\mmet+4~\mathrm{jets}$ & 0.00132 \\
\hline
\hline
\multicolumn{2}{c}{ Longitudinal fraction is $a_0$=1.00}\\
\hline
\ttbar $\goes ZcZc \goes e+\mmet+4~\mathrm{jets}$ & 0.000838 \\
\hline
\ttbar $\goes ZcZc \goes \mu+\mmet+4~\mathrm{jets}$ & 0.00137 \\
\hline
\hline
\end{tabular}
\label{tab:zczc4j_efficiency_a5}
\end{table}

\subsection{Fitting the Top Quark Mass}
\label{fitting_top}
\par We reconstruct the value of $M_{\mathrm{top}}$ for each candidate 
event that contains at least two leptons consistent with a parent
$Z$ boson and at least four jets. The procedure is very similar to
that of the CDF top quark mass measurement~\citep{top_mass_jes}.

\par The value of $M_{\mathrm{top}}$ is calculated by minimizing the
$\chi^2$ distribution, which is based on the assumption that the event
is $p\bar{p}\goes t\bar{t}\goes Z + 4~\mathrm{jets} \goes
\ell\ell + 4~\mathrm{jets}$.  The minimization takes into account
every combination of the jets in the event since we do not know the
true jet-parton assignments. To do so we loop through all possible
permutations and select the one with the lowest $\chi^2$. The top
quark mass
distribution obtained for $t\bar{t}\goes ZcZc\goes\ell\ell +
4~\mathrm{jets}$ decays does not differ significantly from that of
$WbZc$ decay. The exact formula for the $\chi^2$ has the following
structure:
\begin{eqnarray}
&&\chi^2(M_{\mathrm{top}})=\displaystyle\sum_{\ell_1,\ell_2,\mathrm{jets}}\frac{(\hat{Et}_i-Et_i)^2}{\sigma_i^2}+\nonumber\\
&&\sum_{{x,y}}\frac{({\hat{Et}_i}^{uncl}-{{Et}_i}^{uncl})^2}{\sigma_i^2}+\frac{(M(j_1j_2)-M_W)^2}{\Gamma^2_W}+\nonumber\\
&&\frac{(M(l^+l^-) - M_Z)^2}{\Gamma^2_Z}+\frac{(M(W+j)- M_{\mathrm{top}})^2}{\Gamma^2_{\mathrm{top}}}+\nonumber\\
&&\frac{(M(Z+j) - M_{\mathrm{top}})^2}{\Gamma^2_{\mathrm{top}}}.
\end{eqnarray}
The first term contains the fitted transverse energies of the leptons
and four jets within the corresponding experimental resolutions. The
second term includes the $x$- and $y$- components of the unclustered
energy. The expression also contains terms for the reconstructed
masses of the $W$, $Z$, and the two top quarks (i.e. $t\goes
Zc\goes\ell\ell jet$ and $t\goes Wb \goes 3~jets$). The $\chi^2$
function includes all the top-specific corrections of jet energy
scales and energy resolutions used in the single-lepton top quark mass
measurement~\citep{top_mass_jes}.
\par
\begin{figure}[h]
\centering
\includegraphics[angle=0,width=0.45\textwidth]{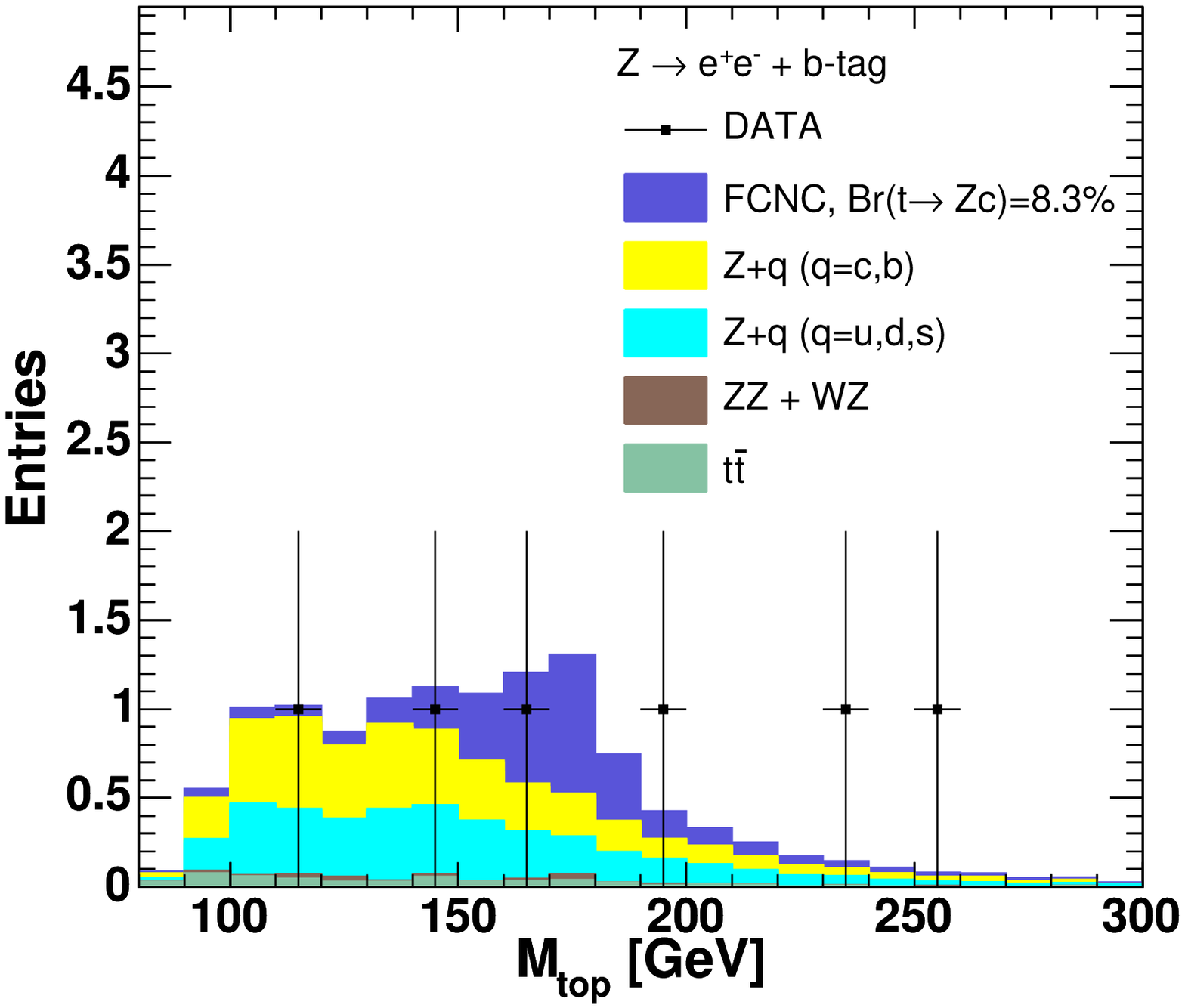}
\includegraphics[angle=0,width=0.45\textwidth]{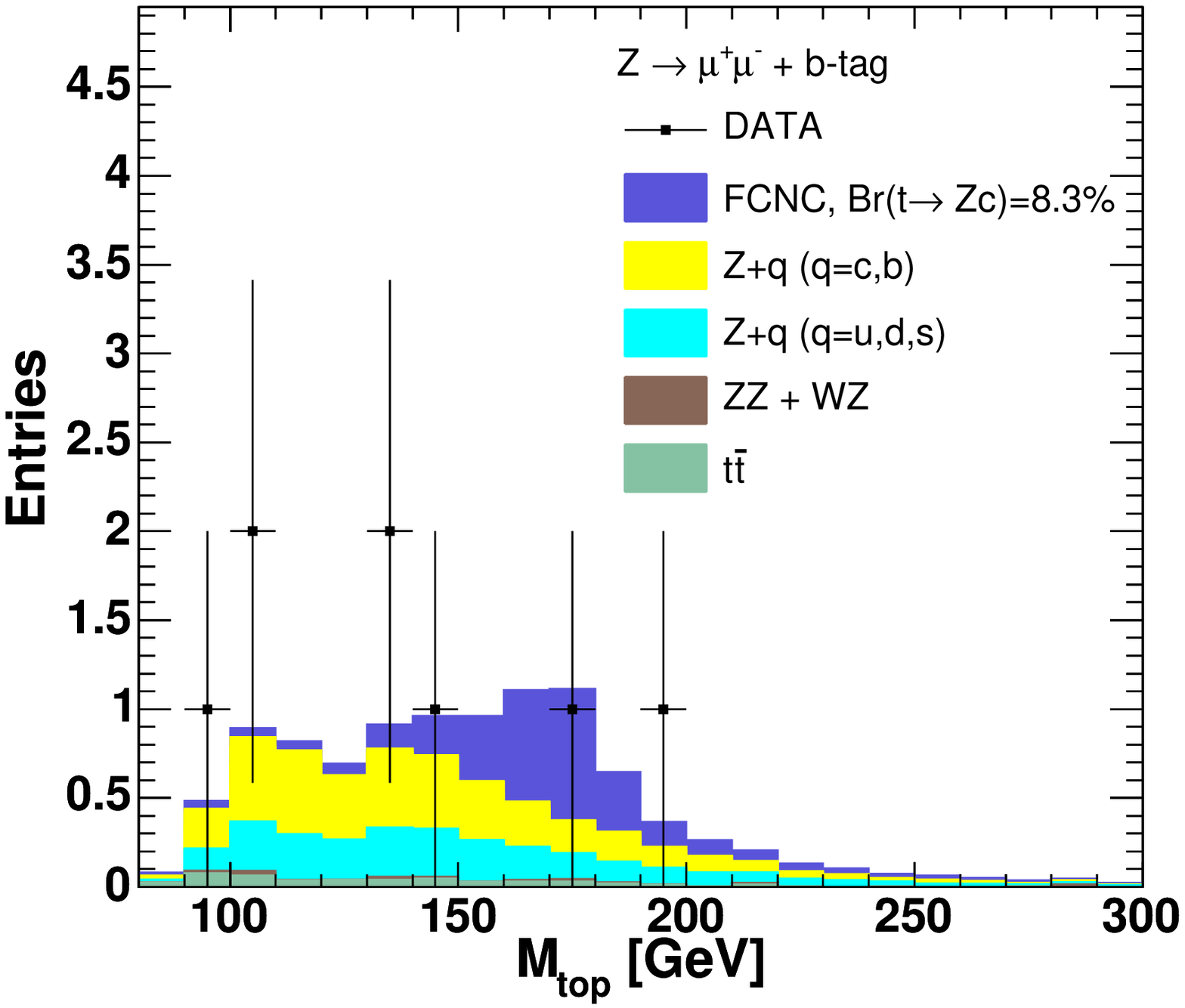}
\caption{The measured distribution (points) in the fitted top quark 
mass in events with a $Z$ and four jets with at least one $b$-tagged
jet, compared to the SM expectations and an FCNC signal (stacked
histogram), for the electron channel (upper figure) and muon channel
(lower figure). The branching fraction for the FCNC signal is taken
from Table~\ref{tab:limits}. The order of stacking in the histograms
is the same as in their legends.}
\label{fig:Z_BTG_TMass}
\end{figure}

 We process the $Z$ + 4 jets events from data and simulation
samples with the same top quark mass fit computer code so that we can
compare the $M_{\mathrm{top}}$ distributions between data, the SM
expectations, and a hypothetical FCNC signal. The comparison is shown
in Fig.~\ref{fig:Z_BTG_TMass}; the data agree well with the SM
background distribution.
In the following section we describe the evaluation of the systematic
uncertainties that go into making this statement quantitative and
setting a limit on a FCNC signal.

\section{Systematic Uncertainties}
\label{systematics}
\par We discuss separately the systematic uncertainties involving the
acceptances and backgrounds in the following two subsections.
\subsection{Systematic Uncertainties on the Acceptances }

\par The uncertainties on the five acceptances used in determining the
limit, $A_{\mathrm{Y}}$, defined in Section~\ref{FCNC_Modeling}, are
summarized in Table \ref{tab:accept_syst}. For each of the
$A_{\mathrm{Y}}$ the effect of uncertainties in the jet energy scale,
initial and final state radiation, lepton identification efficiencies,
parton distribution functions, and the identification (``tagging'') of
bottom quarks and charm quarks have been taken into account.

\begin{table*}[!ht]
\centering
\caption{A summary table of the systematic uncertainties on the
acceptances.  Correlations are taken into account in the calculation
of the limit. The abbreviation ``lept.'' stands for the systematic
uncertainty due to lepton identification and triggering.
}
\begin{tabular}{lccccc}
\hline
Systematic Uncertainty in \% &
 $\frac{\delta(A_{ZZ\goes\ell\ell})}{A_{ZZ\goes\ell\ell}}$ &
 $\frac{\delta(A_{ZW\goes\ell\ell})}{A_{ZW\goes\ell\ell}}$ &
 $\frac{\delta(A_{WZ\goes\ell\mmet})}{A_{WZ\goes\ell\mmet}}$ &
 $\frac{\delta(A_{WW\goes\ell\mmet})}{A_{WW\goes\ell\mmet}}$ &
 $\frac{\delta(A_{ZZ\goes\ell\mmet})}{A_{ZZ\goes\ell\mmet}}$ \\
\hline
\hline
JES & 2.5 & 2.6  & 2.7 & 2.4 & 6.4 \\
\hline
ISR & 0.5& 0.5& 0.5& 0.5 & 0.5 \\
\hline
FSR & 0.6& 0.6& 0.6& 0.6  & 0.6 \\
\hline
PDFs & 0.9 & 0.9 &0.9 & 0.9 & 0.9\\
\hline
HF quark ID & 10.2 & 5.0& 5.0 & 4.1 & 10.6\\ 
\hline
\hline
ID and triggering of electrons & 1.6 & 1.6 & 0.1 & 0.1  & 0.1\\
\hline
 \multicolumn{6}{c}{or} \\
\hline
ID and triggering of muons& 2.8 & 2.8 & 0.9 & 0.9  & 0.9\\
\hline
\hline
Total & 10.6 $\oplus$ lept. & 5.8 $\oplus$ lept.& 5.8 $\oplus$ lept.& 4.9 $\oplus$ lept. & 12.4 $\oplus$ lept.\\
\hline
\end{tabular}
\label{tab:accept_syst}
\end{table*}

\par To estimate the systematic uncertainty from each of these
sources, we vary each of the parameters listed below by one standard
deviation ($\pm\sigma$) and recalculate the acceptances
$A_{\mathrm{Y}}$. The effect of the uncertainty for each of the
sources is correlated among the $A_{\mathrm{Y}}$, and these
correlations are taken into account in the limit-setting procedure.

\par Of the two acceptances that contribute to the signatures
containing two charged leptons, $A_{ZZ\goes\ell\ell}$ and
$A_{ZW\goes\ell\ell}$, the latter dominates as it depends linearly on
the FCNC branching ratio of the $Z$ boson, while the contribution
corresponding to $A_{ZZ\goes\ell\ell}$ enters as the square. In a
similar fashion, the single lepton + \met~signature is dominated by
the SM decay of the top quark pair into $W^+W^-{b\bar{b}}$, with an
acceptance $A_{WW\goes\ell\mmet}$, as there is no FCNC branching
ratio in the rate. The process described by $A_{WZ\goes\ell\mmet}$ is
suppressed by a single factor of the FCNC branching ratio, while that
described by $A_{ZZ\goes\ell\mmet}$ is quadratic, and hence makes a
very small contribution.

\par The largest systematic uncertainties in the dominant processes in
the dilepton and single-lepton modes are the uncertainties in the
efficiency for identifying $b$- and $c$- quarks. For $b$-quarks, we
follow the prescription used previously in CDF studies of the top
quark, and use a systematic uncertainty on the tagging efficiency of
5\%~\citep{mistag_rate}. Similarly, for $c$-quarks, we assign a 15\%
uncertainty~\citep{SECVTX}.

\par The next largest contribution to the systematic uncertainties is
from uncertainties in the calibration of jet
energies~\citep{top_mass_jes}. The systematic uncertainties are
positively correlated for all the $A_{\mathrm{Y}}$. 
%

\par The contributions from lepton identification and trigger
efficiencies are limited by the precision check of the $R$-ratio (see
Section~\ref{R_ratios}). We assume that the reconstruction and the
triggering efficiencies of electrons and muons are not correlated. We
note that acceptances and trigger efficiencies are correlated for
$W\goes\ell\nu$ and $Z\goes\ell\ell$ decays to leptons of the same
flavor. This means that $A_{ZZ\goes\ell\ell}$ would be mis-estimated
by the same percentage as $A_{ZW\goes\ell\ell}$ for leptons of the
same flavor. The same holds true for $A_{WZ\goes\ell\mmet}$,
$A_{WW\goes\ell\mmet}$, and $A_{ZZ\goes\ell\mmet}$.

\par The systematic uncertainties in the $A_{\mathrm{Y}}$ due to
lepton identification and triggering are estimated using deviations
between the measured cross-sections of inclusive $W$'s and $Z$'s, used
in calculating the ratio $R$, from their theoretical values:
\begin{equation}
\frac{\Delta\sigma(Z\goes\ell\ell)}{\sigma(Z\goes\ell\ell)} =
-\frac{\delta(A_{ZZ\goes\ell\ell})}{A_{ZZ\goes\ell\ell}} =
-\frac{\delta(A_{ZW\goes\ell\ell})}{A_{ZW\goes\ell\ell}}
\end{equation}
and
\begin{eqnarray}
\frac{\Delta\sigma(W\goes\ell\nu)}{\sigma(W\goes\ell\nu)} =&& 
-\frac{\delta(A_{WZ\goes\ell\mmet})}{A_{WZ\goes\ell\mmet}} \nonumber\\
=&& -\frac{\delta(A_{WW\goes\ell\mmet})}{A_{WW\goes\ell\mmet}} \nonumber \\
=&& -\frac{\delta(A_{ZZ\goes\ell\mmet})}{A_{ZZ\goes\ell\mmet}}
\end{eqnarray}
The uncertainty on the integrated luminosity does not contribute at
the first order to the measurement of $Br(t\goes Zc)$ since it is
positively correlated between $\sigma(W\goes\ell\nu)$ and
$\sigma(Z\goes\ell\ell)$.

\par The deviation of the measured $R$ ratio is
\begin{eqnarray}
\frac{\Delta R}{R} =&&
\Delta\left(\frac{\sigma(W\goes\ell\nu)}{\sigma(Z\goes\ell\ell)}\right)/
\left(\frac{\sigma(W\goes\ell\nu)}{\sigma(Z\goes\ell\ell)}\right)\nonumber\\
=&& \frac{\Delta\sigma(W\goes\ell\nu)}{\sigma(W\goes\ell\nu)} 
-\frac{\Delta\sigma(Z\goes\ell\ell)}{\sigma(Z\goes\ell\ell)}.
\end{eqnarray}
Therefore, the connection between the deviation in the $R$-ratio and
the uncertainties of the $A_{\mathrm{Y}}$ is the following:
\begin{eqnarray}
\frac{\Delta R}{R} =&&
\delta\left(\frac{A_{ZW\goes\ell\ell}}{A_{WW\goes\ell\mmet}}\right)
/\left(\frac{A_{ZW\goes\ell\ell}}{A_{WW\goes\ell\mmet}}\right)\nonumber\\
=&& \frac{\delta(A_{ZW\goes\ell\ell})}{A_{ZW\goes\ell\ell}}-
\frac{\delta(A_{WW\goes\ell\mmet})}{A_{WW\goes\ell\mmet}}.
\end{eqnarray}
We treat $\frac{\delta(A_{WW\goes\ell\mmet})}{A_{WW\goes\ell\mmet}}$
and $\frac{\delta(A_{ZW\goes\ell\ell})}{A_{ZW\goes\ell\ell}}$ as
negatively correlated as it is the most conservative case. Also this
treatment insures the constraint from the $R$-ratio.


\par Contributions from other sources are significantly smaller than
those from heavy flavor identification and jet-energy scale.  The
effect of initial and final state radiation (ISR and FSR) on
$A_{WW\goes\ell\mmet}$ was studied in Ref.
\citep{top_cross-section}. We expect that FSR will contribute to the
uncertainties in the other three $A_{\mathrm{Y}}$ in the same way
since we require four jets in the final state for all four channels
and the samples are triggered on leptons. The ISR uncertainty should
also contribute identically to the uncertainties of the four
acceptances $A_{\mathrm{Y}}$. The uncertainties are found to be 0.5\%
for ISR and 0.6\% for FSR, assumed to be 100\% correlated across all
$A_{\mathrm{Y}}$.

\par The uncertainties arising from parton distribution functions
(PDF's) can also propagate into the acceptances. However, the dominant
effect of changes in the PDF's is on the production of the $t\bar{t}$
pairs and not on the decay kinematics. The effect of the uncertainties
was also studied in Ref.~\citep{top_cross-section}. The total
uncertainty is 0.9\% and is $100\%$ correlated for the four
$A_{\mathrm{Y}}$.


\subsection{Systematic Uncertainties of the Backgrounds}

\par The sensitivity of this search for a $Z$ boson and a charm quark
coming from top quark decay depends strongly on the understanding of
SM $W$ boson and $Z$ boson production in conjunction with heavy flavor
($W$/$Z$+HF). We summarize the systematic uncertainties of backgrounds
in both the single lepton and di-lepton signatures (the terms
$B_{\ell\nu}$ and $B_{\ell\ell}$ in Equations~\ref{exp_lnu}
and~\ref{exp_ll}) in Table~\ref{tab:bckgr_syst}, and discuss them
below.
\begin{table}[h]
\centering
\caption{The relative systematic uncertainties (\%) on the backgrounds 
for 4-jet semi-leptonic and dilepton final states of \ttbar~pairs.
The contributions from uncertainties in the Monte Carlo modeling and
in the rate of misidentified heavy-flavor jets (mis-tags) are
(conservatively) taken to be correlated in the computation of the
limit. }
\begin{tabular}{lcc}
\hline
Systematic Uncertainty in \% & $\ell\met+4\mathrm{jets}$ & $\ell\ell+4\mathrm{jets}$\\
\hline
\hline
$W$/$Z$+HF+Jets              & 20 & 20 \\
MC Modeling             &     &      \\
\hline
Mis-tags                  & 15 &  15 \\
\hline
$W$/$Z$+HF                   & 2.5 & 8 \\
Normalization            &     &      \\
\hline
\hline
\end{tabular}
\label{tab:bckgr_syst}
\end{table}
\par
\begin{figure}[h]
\centering
\includegraphics[angle=0,width=0.45\textwidth]{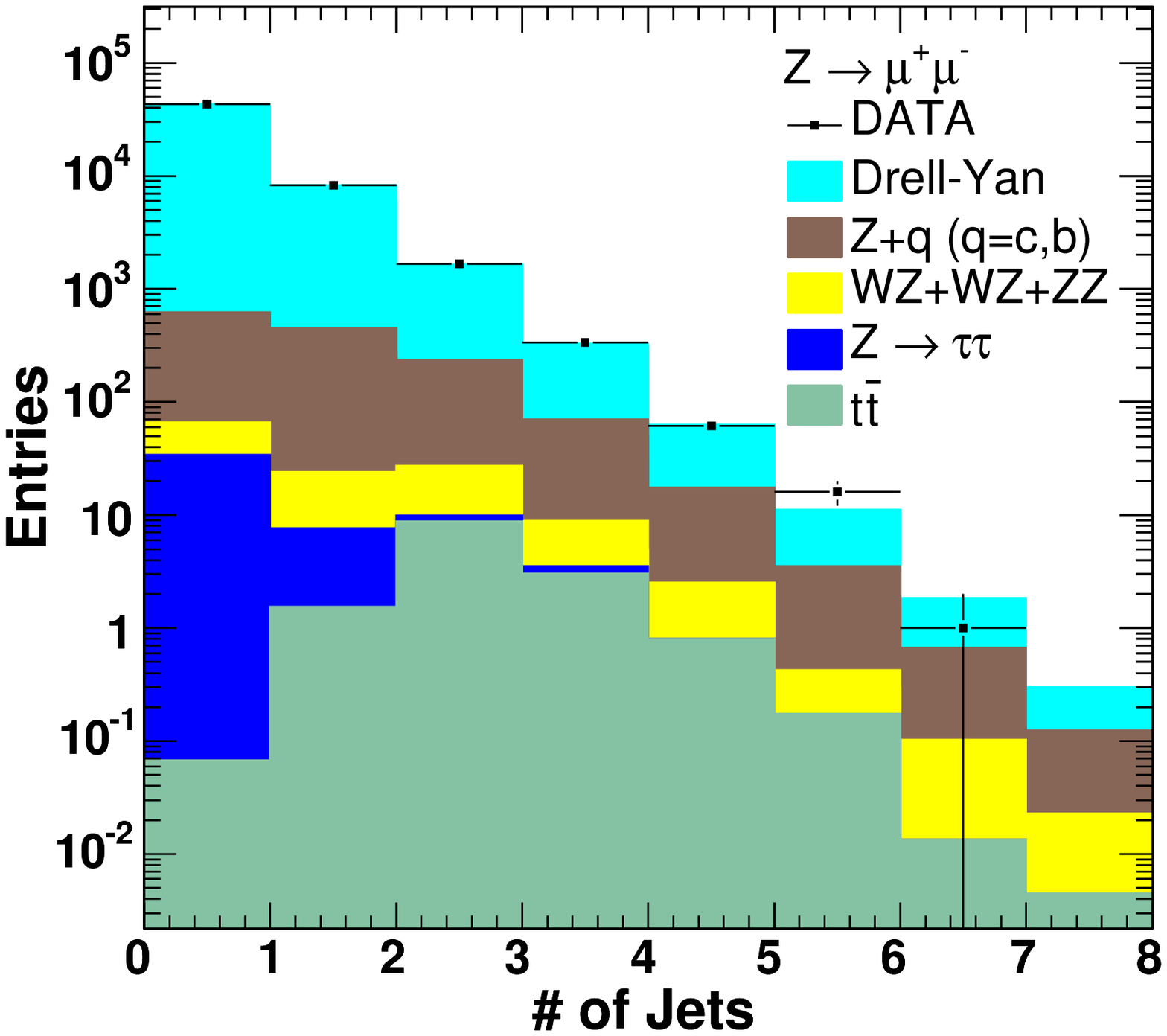}
\includegraphics[angle=0,width=0.45\textwidth]{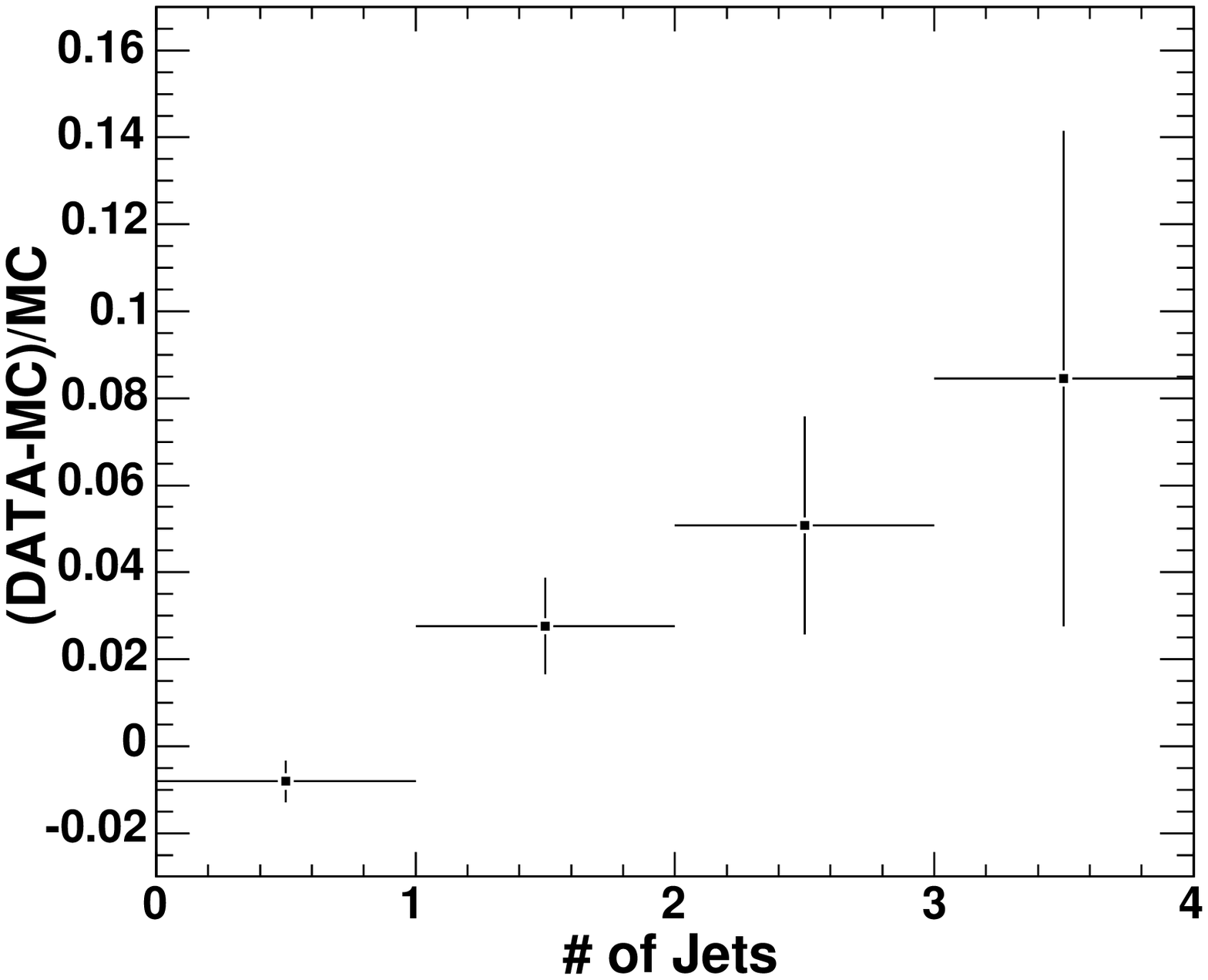}
\caption{The measured distribution (points) in the number of jets in
events with an inclusive decay of \zmumu, compared to SM expectations
(stacked histogram). The order of stacking in the histogram is the
same as in the legend. The $Z$+jets processes (Drell-Yan, $Z$+$b$, and
$Z$+$c$) are modeled with {\sc alpgen}. The lower plot shows the
difference between the data and predictions.}
\label{fig:Z_NJets_Alp}
\end{figure}

 The largest uncertainty in the background comes from modeling the
production of $W$ bosons and $Z$ bosons accompanied by heavy-flavor
and additional jets.  The $Z$+HF and $W$+HF backgrounds are modeled by
{\sc alpgen}~\citep{Alpgen}, and hadronized with {\sc
pythia}~\citep{Pythia}.  The predictions suffer from uncertainties in
the modeling procedure. In particular, the expected number of events
in the $W$/$Z$ + 4 jets category enters directly into the calculation
for the final result. To make an estimate of the uncertainty on the
expected number of $W$/$Z$+HF events, we assume that there is a set or
parameters which allows {\sc alpgen} to model the data precisely. A
deviation from the ``ideal set" can be estimated using inclusive $Z$ +
jets events with jet multiplicity below three. A comparison between
data and {\sc alpgen} simulations is shown in
Fig.~\ref{fig:Z_NJets_Alp}. The observed deviation on the rate of
radiation of one extra jet in the inclusive sample is less than
5\%. We assume independent gluon emission, and so take 10\% as the
estimate of the uncertainty on this {\sc alpgen} prediction for the
radiation of 2 extra jets in the inclusive sample.  However, the
slopes of the N-jet distribution are predicted to be different in the
inclusive and HF samples, with the factors for each additional jet
being 5.0 and 2.7 in the inclusive and $b$-tagged samples,
respectively. The ratio of 5.0 to 2.7 makes a relative difference of
1.85 between radiating an extra jet in inclusive and tagged
samples. We consequently increase the 10\% deviation by a factor of 2
(rounding 1.85 up), to 20\%.
%
\par The sensitivity of the limit to the number of 4-jet $Z$+HF events
was calculated by performing a set of ``pseudo-experiments'' with
different levels of the systematic uncertainties of the backgrounds. The
limit was calculated with this systematic uncertainty set to zero, set
to 20\% (nominal), and set to 40\%. The respective shifts in the limit
are -0.1\%, zero (by construction), and +0.1\%, respectively. The weak
dependence is caused by the measurement technique; we measure a ratio
of top quark events between $Z$ + 4 jets and $W$ + 4 jets final
states. An increase in the number of background events leads to a
decrease in the \ttbar~cross-section measured with $W$ + 4 jets
events. Simultaneously it leads to a decrease in the upper limit on
the number of FCNC signal events in the $Z$ + 4 jets final state.

%
\par The method of predicting misidentified heavy flavor (mis-tags) by
applying a parameterization of the rate for a light-quark jet or gluon
jet being mis-identified as a jet from a charm or bottom quark to jets
in a sample before heavy flavor identification contributes a
significant systematic uncertainty to the background estimates. We
vary the mis-tag probability calculated by the standard CDF algorithm
used in the measurement of the top quark
cross-section~\citep{top_cross-section} jet-by-jet by $\pm 15$ \%
(i.e. a factor of 0.85 or 1.15) to estimate the contribution to the
uncertainty.
\par A smaller contribution to the uncertainty is due to the overall
normalization of the predicted SM boson+HF contribution. The
normalizations of the background distributions from $W$+HF and $Z$+HF
events are treated as independent, and are chosen to match the number
of observed events in the $W$+HF+2 jets and $Z$+HF+2 jets channels,
respectively, as discussed in detail in Sections \ref{ttbar} and
\ref{FCNC_Zjets}. The finite statistics of the 2-jet bin of the data
contributes an uncertainty of 2.5\% to the single lepton and di-lepton
signatures, respectively.
\par The 6\% uncertainty of the measured luminosity affects only 
processes that are normalized absolutely: $WW$, $WZ$, and $ZZ$
production. Consequently, the contribution from the uncertainty of
luminosity to the final result is negligible ($<0.1$\%).
\section{Statistical Evaluation of the Limits on $Br(t\goes Zc)$}
\label{final_limit}
\par At this point we have all the ingredients needed to evaluate
limits on the FCNC branching ratio $Br(t\goes Zc)$. The branching
ratio is evaluated by comparing the numbers of expected and observed
events in two final states, $\ell\ell+4\mathrm{jets}$ and
$\ell\mmet+4\mathrm{jets}$, using Poisson statistics. The numbers of
observed events are denoted as $N_{\ell\nu}$ and $N_{\ell\ell}$ for
final states $\ell\mmet+4\mathrm{jets}$ and $\ell\ell+4\mathrm{jets}$,
respectively, the numbers of expected events are denoted as
$X_{\ell\nu}$ and $X_{\ell\ell}$.

\par To avoid large systematic uncertainties we simultaneously analyze 
two final states from decays of top quark pairs: $p\bar{p}\goes
t\bar{t}\goes ZcWb \goes \ell\ell cjjb$, and $p\bar{p}\goes
t\bar{t}\goes WbWb \goes \ell\mmet bjjb$. This is done by comparing
the number of expected events from SM $t\bar{t}$ decays and SM
backgrounds to the number of observed events in each final state. The
contributions from $t\bar{t}$ decays depend on two numbers: $Br(t\goes
Zc)$ and $N_{t\bar{t}} =
\sigma(p\bar{p}\goes t\bar{t})\int{Ldt}$, where
$\sigma(p\bar{p}\goes t\bar{t})$ is the cross-section of top quark
pair production and $\int{Ldt}$ is the integrated luminosity. We treat
$Br(t\goes Zc)$ and $N_{t\bar{t}}$ as free parameters in the
calculation of the limit on the FCNC branching ratio. The result of
the comparison is presented as a likelihood which is a two-dimensional
function of $Br(t\goes Zc)$ and $N_{t\bar{t}}$. We use the likelihood
distribution to estimate limits on the FCNC branching ratio $Br(t\goes
Zc)$ using a Bayesian approach.

\par For simplicity, let us consider the case in which we observe only
two categories of events: $N_{\ell\nu}$ and $N_{\ell\ell}$, by
applying some set of selection requirements. Later we will show how to
generalize this approach to be used with more categories of selected
events. This is done since we will consider events with electrons and
muons separately and we use a binned distribution of
$M_{\mathrm{top}}$ for $\ell\ell$+4 jets events.

\par We assume that the top quark has only the two decay channels $Wb$
and $Zc$, and so $Br(t\goes Wb) + Br(t\goes Zc) = 1$.  The number of 
expected \ttbar~pairs is
\begin{equation}
N_{t\bar{t}}=\sigma(p\bar{p}\goes t\bar{t})\cdot\displaystyle\int{}{}{Ldt},
\end{equation}
where $\sigma(p\bar{p} \goes t\bar{t})$ can be taken {\it a priori}
since it is independent of any FCNC physics.

\par The expected numbers of events in each of the decay modes are
estimated as follows, where we use the notation $B_Z$ = $Br(t\goes Zc)$:
\begin{equation}
\label{exp_lnu}
X_{\ell\nu} \approx B_{\ell\nu}+N_{t\bar{t}}A_{WW\goes\ell\mmet}
\end{equation}
and
\begin{equation}
\label{exp_ll}
X_{\ell\ell} \approx B_{\ell\ell}+N_{t\bar{t}}A_{ZW\goes\ell\ell}\cdot B_Z.
\end{equation}
The complete formulas are presented in~\citep{equations_expected}. In
the formulas above $B_{\ell\nu}$ and $B_{\ell\ell}$ are non-top SM
contributions (backgrounds) to final states
$\ell\mmet+4\mathrm{jets}$ and $\ell\ell+4\mathrm{jets}$,
respectively; $A_{\mathrm{Y}}$ is acceptance for a decay mode ``Y''
(see Section~\ref{FCNC_Modeling}).

\par The limit on the ratio $Br(t\goes Zc)$ is estimated using
probability density (i.e. likelihood) function defined as:
\begin{equation}
L(B_Z, N_{t\bar{t}}) =  P(N_{\ell\nu}, N_{\ell\ell} | B_Z, N_{t\bar{t}})
\end{equation}
i.e.
\begin{equation}
L( B_Z, N_{t\bar{t}}) = P( N_{\ell\nu}| X_{\ell\nu})P(N_{\ell\ell} | X_{\ell\ell}),
\end{equation}
where
\begin{equation}
P(N|X) = \frac{X^Ne^{-X}}{N!}
\end{equation}
is a Poisson distribution. The likelihood $L(B_Z, N_{t\bar{t}})$ is
defined in the physical region of parameters $ N_{t\bar{t}} \geq 0$
and $0 \leq B_Z \leq 1$. 


\par The complete set of systematic uncertainties is included in the
likelihood function using a Monte Carlo simulation which takes into
account the correlations between the uncertainties.

\par To discriminate the FCNC signal from the expected SM background,
we use the distribution in the reconstructed top-quark mass,
$M_{\mathrm{top}}$, for $Z$+4 jets events. Events from the signal
process should form a distinguishable peak at the top quark mass. We
combine probabilities for each bin of the reconstructed top-quark mass
distribution
\begin{equation}
\prod_i P(N^i_{\ell\ell} | X^i_{\ell\ell}),
\end{equation}
where the index $i$ refers to the $i$'th bin of the distribution in
the top mass. This requires calculating the acceptances $A_{ZZ\goes
\ell\ell}^{i}$ and $A_{ZW\goes\ell\ell}^{i}$ for each bin of the
reconstructed top quark mass histogram.

\par We note that the electron and muon decay modes of the top quarks
are treated separately up to this point of the analysis in order to
better understand the systematics of both. The two channels are then
included together in the final likelihood function
$L(B_Z,N_{t\bar{t}})$.

\par The likelihood function is used to construct a posterior
probability density $P(B_Z|$DATA), where DATA refers to the numbers of
observed events, $N_{\ell\nu}$ and $N_{\ell\ell}^i$, in the electron
and muon channels ($\ell=e$ or $\ell=\mu$). The posterior probability
density function is converted into a limit on the FCNC branching ratio
$Br(t\goes Zc)$ using a Bayesian approach.

\subsection{Numerical Computation of the Likelihood Distribution
Function $L(B_Z, N_{t\bar{t}})$}

\par
\begin{figure}[ht]
\centering
\includegraphics[angle=0,width=0.5\textwidth]{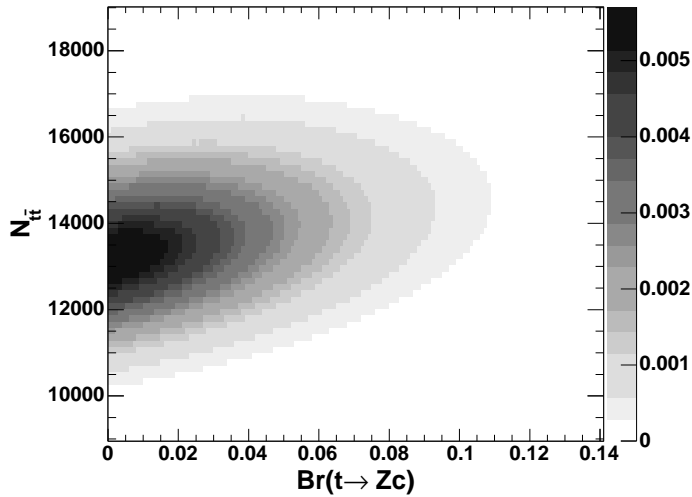}
\caption{ The likelihood distribution 
$L(B_Z, N_{t\bar{t}})$ calculated as a function of $N_{t\bar{t}}$ and
$B_Z$ = $Br(t\goes Zc$). The distribution is for FCNC decays of
$t\goes Zc$ with 100\% longitudinally polarized $Z$ bosons.}
\label{fig:lim2}
\end{figure}

 The observed distribution of the likelihood (computed for
$t\goes Zc$ decays where the $Z$ bosons are 100\% longitudinally
polarized) is presented in Fig.~\ref{fig:lim2}.
\par A likelihood distribution is calculated for each given value of
helicity of the $t\goes Zc$ coupling since the acceptances
$A_{\mathrm{Y}}$ vary for different structures of the FCNC coupling.

\subsection{Computation of the Posterior 
	    $P(Br(t\goes Zc)|\mathrm{DATA)}$ }

\par The posterior probability density functions  $P(B_Z|$DATA)
are computed from the likelihood functions $L(B_Z, N_{t\bar{t}})$
using a Bayesian approach as follows:
\begin{equation}
P(\mathrm{DATA}|B_Z) =\displaystyle\int_{0}^{\infty}L(B_Z, N_{t\bar{t}})\cdot\pi_0(N_{t\bar{t}}) dN_{t\bar{t}}
\end{equation}
\begin{equation}
P(B_Z|\mathrm{DATA}) = \frac{P(\mathrm{DATA}|B_Z)\cdot\pi_1(B_Z)}{\displaystyle\int_{0}^{1}P(\mathrm{DATA}|B_Z)\cdot\pi_1(B_Z)dB_Z},
\end{equation}
where $\pi_0(N_{t\bar{t}})$ is the {\it a priori} probability density
function of $N_{t\bar{t}}$ and $\pi_1(B_Z)$ is the {\it a priori}
distribution of $B_Z$ which is taken to be flat in the physical region
(it is 1.0 for \mbox{$0 \leq B_Z \leq 1$} and zero everywhere
else). The distribution of $\pi_0(N_{t\bar{t}})$ represents the prior
knowledge of the top pair production cross-section,
$\sigma(p\bar{p}\goes t\bar{t})$.

\par We consider two choices of the $\pi_0(N_{t\bar{t}})$ prior
distribution: flat and Gaussian. The flat distribution does not
contain any information regarding the theoretical predictions of
$\sigma(p\bar{p}\goes t\bar{t})$. It is just a constant. The Gaussian
distribution is derived using the theoretical estimates of the top
pair production cross-section $\sigma(p\bar{p}\goes
t\bar{t})$~\citep{sigttbar} and the integrated luminosity. The
theoretical estimate of the top pair production cross-section is
presented as a function of top quark mass $M_{\mathrm{top}}$. The
measured top-quark mass is 170.9 $\pm$ 1.8 \GeV~\citep{top_mass}. The
luminosity is \lum, with an uncertainty of 6\%. The Gaussian prior
allows us to take into account the theoretical FCNC-independent
knowledge of $\sigma(p\bar{p}\goes t\bar{t})$.

\par
\begin{figure}[ht]
\centering
\includegraphics[angle=0,width=0.5\textwidth]{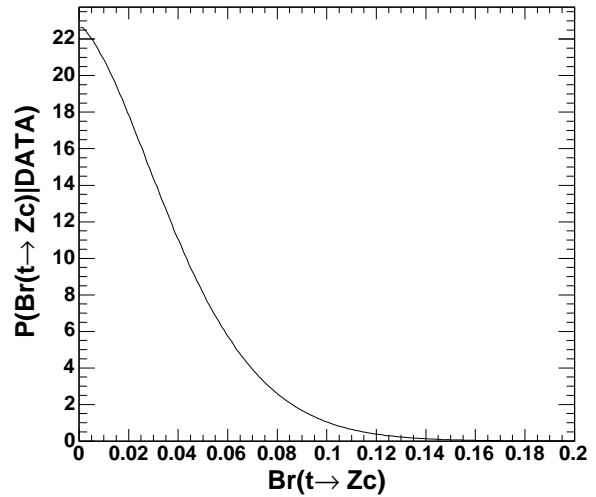}
\caption{The distribution for $P(Br(t\goes Zc)|$DATA), calculated for 
100\% longitudinally polarized $Z$ bosons using Guassian prior.}
\label{fig:1dcont}
\end{figure}

 The distribution for $P(B_Z|$DATA), calculated for 100\%
longitudinally polarized $Z$ bosons, is shown in
Fig.~\ref{fig:1dcont}.

\subsection{Computation of the Upper Limits on $Br(t\goes Zc)$}
\par We use the posterior function $P(B_Z|$DATA) to calculate the
upper limit $B_Z^{lim}$ on $Br(t\goes Zc)$ (i.e. $B_Z$) by solving the
equation:
\begin{equation}
\beta = \displaystyle \int_0^{B_Z^{lim}} P(B_Z|\mathrm{DATA}) dB_Z,
\end{equation}
where $\beta$ is 0.95 (95\% C.L). The upper limits versus
the helicity of the $Z$ boson are summarized in Table~\ref{tab:limits}.

\par We perform statistical cross-checks of the measured upper limits
using pseudo-experiments. The pseudo-experiments are generated
randomly assuming that there is no contribution from FCNC processes,
i.e. by setting $Br(t\goes Zc)=0$. The expected upper limit for 100\%
longitudinally polarized $Z$'s on $Br(t\goes Zc)$ is 8.7$\pm$2.6\%, 
consistent with the observed limit of 8.3\%. 

\section{Conclusions and Results}
\label{conclusions}

\par Taking into account systematic uncertainties on Monte Carlo
simulations, $b$-tagging, mis-tag modeling, and lepton identification,
we find an upper limit at 95\% C.L. on the branching ratio of $t\goes
Zc$ of 8.3\% for FCNC decays where the $Z$ bosons are 100\%
longitudinally polarized. The result is primarily statistics-limited.
It can be significantly improved with more data if the number of
$Z$+4 jets events is high enough to do a shape analysis of the
top quark mass distribution.



\par To be assumption-independent we parametrize the limit on
$Br(t\goes Zc)$ as a function of the fraction of longitudinally
polarized $Z$ bosons. The parametrization allows us to cover the full
range of all possible helicity structures of the $t\goes Zc$
vertex. The upper limits are calculated at 95\% C.L. for five
fractions of longitudinally polarized $Z$'s using \lum~of data. The
results are presented in Table \ref{tab:limits} for both the Gaussian
and the flat priors. The limits vary between 8.3 and 9.0\% for the
Gaussian prior depending on the polarization of the $Z$ boson and are
about 1\% less restrictive for the flat prior.
\begin{table}[h]
\centering
\caption{ The upper limits on the FCNC branching ratio $Br(t\goes Zc)$ 
in \% as a function of the longitudinal fraction of the $Z$ bosons in
the FCNC coupling ($t\goes Zc$) at 95\% CL. The limits labeled
Gaussian prior use as input the theoretical cross-section of
$\sigma(p\bar{p}\goes t\bar{t})$; the limits labeled Flat prior are
theory-independent. }
\begin{tabular}{lccccc}
\hline
\hline
Longitudinal Fraction & 0.00  & 0.25 & 0.50  & 0.75 & 1.0 \\
\hline
Gaussian prior & 9.0\% & 8.8\%  & 8.6\%  &  8.5\% & 8.3\% \\
\hline
Flat prior & 10.2\% & 10.0\%  & 9.7\%  &  9.5\% & 9.2\% \\
\hline
\hline
\end{tabular}
\label{tab:limits}
\end{table}

\section{Acknowledgments}
\label{acknowledgments}
\par We thank the Fermilab staff and the technical staffs of the
participating institutions for their vital contributions. We thank
Mary Heintz for unfailing computer support. We are grateful to Michel
Herquet for explaining to us how to incorporate the FCNC couplings
into {\sc madgraph} code, and then installing the model in {\sc
madgraph}.

\par This work was supported by the U.S. Department of Energy and
National Science Foundation; the Italian Istituto Nazionale di Fisica
Nucleare; the Ministry of Education, Culture, Sports, Science and
Technology of Japan; the Natural Sciences and Engineering Research
Council of Canada; the National Science Council of the Republic of
China; the Swiss National Science Foundation; the A.P. Sloan
Foundation; the Bundesministerium f\"ur Bildung und Forschung,
Germany; the Korean Science and Engineering Foundation and the Korean
Research Foundation; the Science and Technology Facilities Council and
the Royal Society, UK; the Institut National de Physique Nucleaire et
Physique des Particules/CNRS; the Russian Foundation for Basic
Research; the Ministerio de Ciencia e Innovaci\'{o}n, and Programa
Consolider-Ingenio 2010, Spain; the Slovak R\&D Agency; and the
Academy of Finland.


%
\bibliographystyle{apsrev}
\bibliography{prd}

\begin{thebibliography}{61}
\bibitem[{EtP()}]{EtPt}
\bibinfo{note}{Transverse momentum and energy are defined as $\Pt =
  p\sin\theta$ and $\Et = E\sin\theta$, respectively. Missing $\rm E_T$
  ($\metvec$) is defined by $\metvec = -\sum_{i} E_T^i \hat{n}_i$, where i is
  the calorimeter tower number for $|\eta| <$ 3.6 (see Ref.~\cite{CDF_coo}),
  and $\hat{n}_i$ is a unit vector perpendicular to the beam axis and pointing
  at the i$^{th}$ tower. We correct $\metvec$ for jets, photons, electrons and
  muons. We define the magnitude $\met=|\metvec|$. We use the convention that
  ``momentum'' refers to $pc$ and ``mass'' to $mc^2$.}
\bibitem[{CDF({\natexlab{b}})}]{CDF_coo}
\bibinfo{note}{The CDF coordinate system is cylindrical and right-handed, with
  the $x$ axis horizontal and out of the Tevatron ring, the $y$ axis up, the
  $z$ axis along the proton beam, and $r=\sqrt{x^2+y^2}$; $\phi$ is the
  azimuthal angle. The pseudorapidity $\eta\equiv-{\rm ln}(\tan (\theta/2))$,
  where $\theta$ is the polar angle; the $\eta$ regions for detector components
  are defined with respect to the center of the detector. The central region is
  defined as $|\eta|<1$.}
\bibitem[{ele()}]{electron_track_quality}
\bibinfo{note}{For central electrons at least 5 hits in each of 3 axial and 2
  stereo layers of the COT are required.}
\bibitem[{had()}]{hadoem}
\bibinfo{note}{The fraction of electromagnetic energy allowed to leak into the
  hadron compartment $\rm E_{had}/E_{em}$ must be less than $\rm
  0.055\plus0.00045\times E_{em}(\GeV)$ for central electrons, less than 0.05
  for electrons in the end-plug calorimeters, less than max[0.125, $\rm
  0.055\plus0.00045\times E_{em}(\GeV)$] for photons.}
\bibitem[{muo({\natexlab{a}})}]{muon_track_quality}
\bibinfo{note}{For tight muons at least 5 hits in each of 3 axial and 3 stereo
  layers of the COT are required; for loose muons with a matching muon stub
  this is relaxed to 3 axial and 2 stereo. Loose muons without a matching stub
  have an additional cut on the $\chi ^2$ for the fit to the track.}
\bibitem[{muo({\natexlab{b}})}]{muon_cal_cuts}
\bibinfo{note}{The energy deposited in the calorimeter tower traversed by the
  muon must be less than $2+\mathrm{max}(0,0.0115\times(p-100))~\GeV $ in the
  electromagnetic compartment and less than
  $6+\mathrm{max}(0,0.028\times(p-100))~\GeV$ in the hadronic compartment.}
\bibitem[{muo({\natexlab{c}})}]{muon_stub_matching}
\bibinfo{note}{The muon ``stub'' in the muon systems must be within 7, 5, and 6
  cm of the extrapolated COT track position, in the CMU, CMP, and CMX muon
  systems, respectively.}
\bibitem[{lep()}]{lepton_scale_factors}
\bibinfo{note}{The maximum correction to the identification efficiency for
  central electrons is 1.7\%, for ``plug'' electrons 6.3\%, and for central
  muons is 7.4\%.}
\bibitem[{tra()}]{transverse_mass}
\bibinfo{note}{The transverse mass is defined as $M_{\mathrm{trans}}(\ell\nu)$
  = $\sqrt{2p_{\rm T}(\ell)p_{\rm T}(\nu)(1-\cos{(\phi(\ell)-\phi(\nu))})}$,
  where \pt$(X)$ is the transverse momentum of $X$ ($X$ is $\ell$ or $\nu$) and
  $\phi(X)$ is the azimuthal angle of $X$.}
\bibitem[{alp()}]{alpgen_matching}
\bibinfo{note}{We use Version 2.10 of {\sc alpgen}.}
\bibitem[{mon()}]{monte_carlo_details}
\bibinfo{note}{{\sc pythia} v6.216 with a $M(\gamma^*/Z)$ $>$ 30 GeV cut, Tune
  A, and the ``Willis Sakumoto'' corrections
  applied~\cite{Willis_tuned_Pythia}. The $Z$ + HF samples are produced with
  {\sc alpgen} v2.10-prime (which has built-in MLM
  matching~\cite{MLM_matching}) and showered with {\sc pythia} v 6.326. The
  both generators used CTEQ5L parametrization of the parton distribution
  functions.}
\bibitem[{top()}]{top_76pb}
\bibinfo{note}{The theoretical estimate of the top pair production
  cross-section is based on \citep{sigttbar}~ using a top-quark mass of 170.9
  $\pm$ 1.8 GeV \citep{top_mass}.}
\bibitem[{no_()}]{no_signal}
\bibinfo{note}{The scale factor for W+HF processes ($W+b\bar{b}$, $W+c\bar{c}$,
  and $W+c$) does not depend strongly on the number of $t\bar{t}$ events since
  the $t\bar{t}\goes WbZc$ and $t\bar{t}\goes WbWb$ modes contribute mostly to
  a final state with a W and three or four jets in the final state.}
\bibitem[{fou()}]{four_jet_fit}
\bibinfo{note}{The fit for $W\goes\mu\nu+4$ jets with at least one jet b-tagged
  returns zero events with a large uncertainty, due to the small statistics of
  the 4-jet sample. We consequently use the value of 2.6\% obtained from the
  fit for inclusive $W\goes\mu\nu+4$ jets, consistent with the fit.}
\bibitem[{equ()}]{equations_expected}
\bibinfo{note}{The complete formulas for the expected numbers of events are:
  \begin{eqnarray} \label{exp_lnu_mybib} X_{\ell\nu}=&&B_{\ell\nu} +
  N_{t\bar{t}}\cdot\{(1-B_Z)^2\cdot A_{WW\rightarrow\ell~\mmet}+\nonumber \\
  &&B_Z(1-B_Z)\cdot(A_{WZ\rightarrow\ell~\mmet})+\nonumber\\ &&B_Z^2\cdot
  A_{ZZ\rightarrow\ell~\mmet}\} \end{eqnarray} and \begin{eqnarray}
  \label{exp_ll_mybib} X_{\ell\ell}=&&B_{\ell\ell} +
  N_{t\bar{t}}\cdot\{A_{ZW\rightarrow\ell\ell}\cdot B_Z+\nonumber \\
  &&(A_{ZZ\rightarrow\ell\ell}-A_{ZW\rightarrow\ell\ell})\cdot B_Z^2\}.
  \end{eqnarray}}.
\end{thebibliography}
\end{document}